\documentclass{aa}

\usepackage{graphicx}
\usepackage{txfonts}

\usepackage{color}
\usepackage{ulem}
\usepackage{bm}
\usepackage{natbib}
\usepackage{gensymb}
\usepackage{amsmath}

\usepackage{hyperref}
\usepackage{url}

\usepackage{comment}

\definecolor{darkgreen}{rgb}{0.1,0.5,0.1}

\newcommand{\editt}[1]{\textcolor{black}{#1}}
\newcommand{\mat}[1]{\bm{\mathrm{#1}}}

\newcommand{\ShapePipe}{\textsc{ShapePipe}}

\renewcommand{\sout}[1]{\unskip}

\begin{document}

   \title{ShapePipe: a new shape measurement pipeline and weak-lensing application to UNIONS/CFIS data}

   \subtitle{}

   \author{Axel Guinot \inst{1, 2}\fnmsep\thanks{axel.guinot.astro@gmail.com} \and
          Martin Kilbinger \inst{1, 3} \and
          Samuel Farrens \inst{1} \and
          Austin Peel\inst{1, 4} \and
          Arnau Pujol \inst{1,5,6} \and
          Morgan Schmitz \inst{1,7} \and
          Jean-Luc Starck \inst{1}
          \and
          Thomas Erben\inst{11} \and
          Raphael Gavazzi\inst{3, 17} \and
          Stephen Gwyn\inst{16} \and
          Michael J. Hudson\inst{8, 9, 10} \and
          Hendrik Hildebrandt\inst{13} \and
          Tobias Liaudat  \inst{1} \and
          Lance Miller \inst{15} \and
          Isaac Spitzer \inst{8, 10} \and
          Ludovic Van Waerbeke\inst{14}
          \and
          Jean-Charles Cuillandre\inst{1} \and
          S\'ebastien Fabbro\inst{12} \and
          Alan McConnachie \inst{12} \and Yannick Mellier\inst{3}
          }

   \institute{%
    AIM, CEA, CNRS, Universit\'e Paris-Saclay, Universit\'e de Paris, F-91191 Gif-sur-Yvette, France
    \and
    Université de Paris, CNRS, Astroparticule et Cosmologie, F-75013 Paris, France
   \and
    Institut d'Astrophysique de Paris, UMR7095 CNRS, Universit\'e Pierre \& Marie Curie, 98 bis boulevard Arago, F-75014 Paris, France
   \and
   Institute of Physics, Laboratory of Astrophysics, Ecole Polytechnique Fédérale de Lausanne (EPFL), Observatoire de Sauverny, 1290 Versoix, Switzerland
   \and
   Institut d’Estudis Espacials de Catalunya (IEEC), E-08034 Barcelona, Spain
   \and
   Institute of Space Sciences (ICE, CSIC), E-08193 Barcelona, Spain
   \and
    Department of Astrophysical Sciences, Princeton University, 4 Ivy Ln., Princeton, NJ08544, USA
    \and
    Department of Physics and Astronomy, University of Waterloo, Waterloo, ON, N2L 3G1, Canada
    \and
    Waterloo Centre for Astrophysics, Waterloo, ON, N2L 3G1, Canada
    \and
    Perimeter Institute for Theoretical Physics, 31 Caroline St. N., Waterloo, ON, N2L 2Y5, Canada
    \and
    Argelander Institute for Astronomy, University of Bonn, Auf dem H\"ugel 71, 53121 Bonn, Germany
    \and
    NRC Herzberg Astronomy and Astrophysics, 5071 West Saanich Road, Victoria, BC V9E 2E7, Canada
    \and
    Ruhr University Bochum, Faculty of Physics and Astronomy, Astronomical Institute (AIRUB), German Centre for Cosmological Lensing, 44780 Bochum, Germany
    \and
    Department of Physics and Astronomy, University of British Columbia, 6224 Agricultural Road, Vancouver, V6T 1Z1, BC, Canada
    \and
    Department of Physics, University of Oxford, Denys Wilkinson Building, Keble Road, Oxford OX1 3RH, UK
    \and
    Canadian Astronomy Data Centre, Herzberg Astronomy and Astrophysics, National Research Council, 5071 West Saanich Rd, Victoria BC, V9E 2E7
    \and
    Institute of Astronomy, University of Cambridge, Madingley Road, Cambridge CB30HA, UK
   }

   \date{Received; accepted}

 
  \abstract
   {The Ultraviolet Near-Infrared Optical Northern Survey (UNIONS) is an ongoing collaboration that will provide the largest deep photometric survey of the Northern sky in four optical bands to date. As part of this collaboration, the Canada-France Imaging Survey (CFIS) is taking $r$-band data with an average seeing of $0.65~\mathrm{arcsec}$, which is complete to magnitude 24.5 and thus ideal for weak-lensing studies.}
   {We perform the first weak-lensing analysis of CFIS $r$-band data over an area spanning $1\,700$ deg$^2$ of the sky. We create a catalogue with measured shapes for $40$ million galaxies, corresponding to an effective density of $6.8$ galaxies per square arcminute, and demonstrate a low level of systematic biases. This work serves as the basis for further cosmological studies using the full UNIONS survey of $4\,800$ deg$^2$ when completed.}
   {Here we present \ShapePipe, a newly developed weak-lensing pipeline. This pipeline makes use of state-of-the-art methods such as \textsc{Ngmix} for accurate galaxy shape measurement. Shear calibration is performed with metacalibration. We carry out extensive validation tests on the Point Spread Function (PSF), and on the galaxy shapes. In addition, we create realistic image simulations to validate the estimated shear.}
   {We quantify the PSF model accuracy and show that the level of systematics is low as measured by the PSF residuals. Their effect on the shear two-point correlation function is sub-dominant compared to the cosmological contribution on angular scales $<100\arcmin$. The additive shear bias is below $5\times10^{-4}$, and the residual multiplicative shear bias is at most $10^{-3}$ as measured on image simulations. Using COSEBIs we show that there are no significant B-modes present in second-order shear statistics. We present convergence maps and see clear correlations of the E-mode with known cluster positions. We measure the stacked tangential shear profile around Planck clusters at a significance higher than $4\sigma$.}
   {}


    \maketitle
%

\section{Introduction}

    Weak gravitational lensing, the apparent distortion of the shapes of galaxies by foreground matter, is today one of the primary probes of cosmology. The ability to trace the total matter distribution, including dark matter, makes weak-lensing an indispensable tool in the modern era of precision cosmology (see {\it e.g.} \cite{K15} or \cite{2017arXiv171003235M} for reviews). Weak-lensing distortions of galaxy images induced by large-scale structure are very small and prone to a number of systematic errors. In addition, for the cosmological interpretation of the measured gravitational shear, photometric redshifts of the lensed galaxies need to be known to a high precision. For these reasons, weak-lensing studies require a very large observed area, high image quality in multiple bands for photometric redshifts, and a significant depth.
    
    Previous experiments have studied weak-lensing in great detail, such as the Canada-France-Hawai'i Telescope Legacy Survey \citep[CFHTLS;][]{CFHTLenS-data}, the Kilo-Degree Survey \citep[KiDS;][]{Kuijken_2019}, the Hyper Suprime Cam survey \citep[HSC;][]{2018PASJ...70S..25M}, or the Dark Energy Survey \citep[DES;][]{gatti2020dark}.
    
    In this paper we focus on the ongoing Ultra-violet Near-Infrared Optical Northern Survey (UNIONS). UNIONS is a collaboration between the Canada-France Imaging Survey (CFIS), Pan-STARRS \citep{panstarrs_ps1}, and WISHES (Wide Imaging with Subaru HSC of the Euclid Sky). It aims to provide the largest ($4\,800~\mathrm{deg}^{2}$) multi-band optical photometric survey of the Northern hemisphere. CFIS provides $r$-band images observed with the Canada-France Hawai'i Telescope (CFHT) with excellent image quality, ideal for weak-lensing purposes (the average seeing is $\approx0.65$~arcsec in the $r$-band). CFHT has a proven track record of providing images for state-of-the-art weak-lensing studies. Wide surveys with CFHT started with CFHTLS/CFHTLenS followed by RCSLenS \citep{Hildebrandt_2016} and now CFIS. In addition to the large area and small seeing, CFIS area overlaps with very wide spectroscopic surveys such as SDSS-BOSS \citep{2011AJ....142...72E}, eBOSS \citep{2016AJ....151...44D}, and soon DESI \citep{DESI_science_17}. This combination of surveys provides a unique data set for weak-lensing studies.
     
    Central to all weak-lensing analyses is a robust and efficient data processing pipeline. To extract the weak-lensing signal from the distorted shapes of the galaxies one has to be particularly meticulous through the entire chain of processing. The main piece of such a pipeline is the shape measurement algorithm used to capture the shear signal from the noisy, pixelised, and blurred images of galaxies. This signal is subject to a number of systematic errors. Every step requires specific calibration and validation to reach the level of precision required for the cosmological analysis.
    
    In this paper we present a new pipeline architecture designed to handle the large area of a Stage-III \citep{DETF} survey. The pipeline presents a balance of well-established methods, and newly developed algorithms. Our goal is to develop a framework that is capable of handling CFIS data, but is flexible enough to allow for the evolution of the current methods and the addition of new ones in the future.
    
    This paper is organised as follows.
    In Sect.~\ref{sec:data} we present the UNIONS survey with a focus on the CFIS $r$-band, which was used for the shape measurement presented here. In this section we also detail how we estimate the redshift distribution for this first study. In Sect.~\ref{sec:pipeline} we introduce the modular design of our pipeline. In Sect.~\ref{sec:psf_section} we discuss the modelling of the point spread function (PSF), a key part for any weak-lensing pipeline. In Sect.~\ref{sec:shape_measurement} we present the multi-epoch shape measurement method. Sect.~\ref{sec:diagnostics} shows the numerous diagnostics we carry out to validate our measurements on the data. We also perform tests on simulated images to validate our implementation of the shape measurement. These tests are presented in Sect.~\ref{sec:simulation}. Finally, we show our scientific results in Sect.~\ref{sec:science} before presenting our conclusions in Sect.~\ref{sec:conclusion}.


\section{Data}
\label{sec:data}
    
    CFIS is a large imaging survey observing the Northern hemisphere with the wide-field imager \textsc{MegaCAM} on CFHT (pixel scale of 0.187 arcsec). CFIS started observations in 2017 and has (by early 2021) reached a coverage of around $3\,000$ deg$^2$, $60\%$ of the planned final area of $4\,800~\mathrm{deg}^{2}$. The survey will reach completion by 2025. CFIS takes deep images in the $r$-band ($640~\mathrm{nm}$) and u-band ($355~\mathrm{nm}$). It takes advantage of the excellent sky quality of Mauna Kea, with an average seeing of $0.65~\mathrm{arcsec}$ in the $r$-band. These conditions make CFIS ideally suited for weak-lensing studies.
    
    In 2018 the UNIONS (Ultraviolet Infra-Red Northern Sky) survey collaboration was created to gather in a single scientific group the various multi-band surveys covering the planned Euclid footprint in the Northern hemisphere. CFIS will provide the r- and u-bands, while Pan-STARRS will observe the i- and z-band, and Subaru HSC will complement the z-band through WISHES.

    \subsection{The UNIONS/CFIS survey}

      \subsubsection{Observation strategy}
      
          CFIS single-exposure images are taken with a large dither between each exposure, which is around one third of the focal plane, or $0.33$ deg. This survey strategy was chosen to maximise the total area covered. Three exposures (of $\sim$200 seconds each) are necessary to reach the planned depth of $r=24.1$ at SNR=10 for extended sources observed in the stacked images. The decision was made to have an exposure time that varies (between 100s and 300s) according to observing conditions (sky brightness, image quality, sensitivity of the system telescope+camera) to achieve a constant magnitude depth across the survey footprint \citep{observing_strategy}.

    \subsection{Pre-processing steps}
    
        The pre-processing of the data is a key step in the processing chain. For CFIS, CFHT images are calibrated using the  MegaPipe pipeline \citep{MegaPipe_paper}. An astrometric calibration within 20~mas was achieved using the Gaia DR2 observations \citep{gaia_dr2} of 1.7 billions stars. The photometric calibration relies on observations from the Pan-STARRS PS1 survey \citep{panstarrs_ps1} providing a photometric solution as good as 1 milli-magnitude in the $r$-band internally (camera field-of-view), and 4 milli-magnitudes in absolute with respect to an all-sky reference. Both of these steps are important for shape measurement as the gravitational lensing signal is extremely sensitive to astrometric and photometric calibrations. The level of calibration achieved by CFIS can reduce systematic effects \citep{2017arXiv171003235M}.
        
        After the single exposures have been calibrated, stacked images are created. This step allows one to increase the signal-to-noise ratio by combining several images together. For CFIS, the images are combined with \textsc{SWARP}\footnote{https://github.com/astromatic/swarp} using a weighted average. This method has the advantage of preserving point-source-like objects, but it does not behave as well as a median average regarding outliers. This choice was motivated by the small number of single exposures (three on average), which is not enough to properly remove time-dependent artefacts even with a median stacking.
        
    \subsection{Galaxy redshift distribution}
    \label{sec:redshift_distribution}
    
        As mentioned above, the UNIONS survey will be composed of several photometric bands, making it possible to compute photometric redshifts (photo-$z$'s) for all observed galaxies. However, at present time only the $r$-band data have reached a sufficiently large area and depth to obtain reliable photo-$z$'s. For our science analysis (Sect.~\ref{sec:science}) we need an estimate for the redshift distribution $N(z)$ of our weak-lensing galaxies over the processed area.
        
\begin{figure}
            \centering
            \includegraphics[width=1.\linewidth]{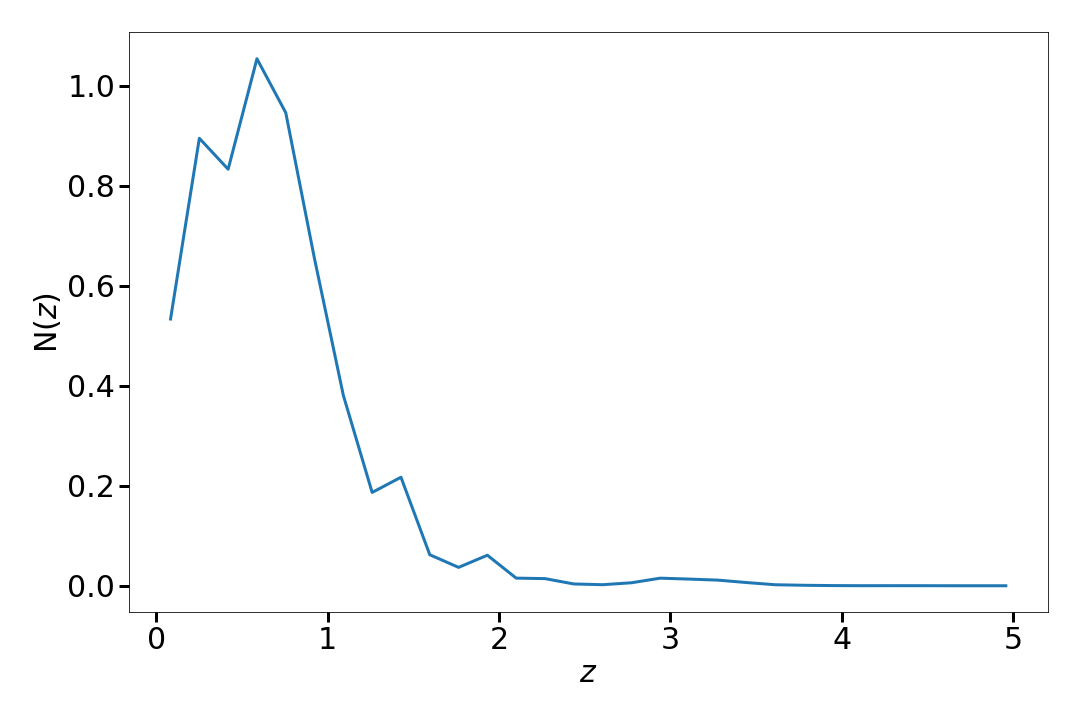}
            \caption{Redshift distribution inferred from matching CFIS with CFHTLenS \texttt{Z\_BEST} over the CFHTLS-W3 area.}
            \label{fig:n_of_z}
        \end{figure}

        Here, we make use of the overlapping area with the CFHTLS-W3 field analysed by the CFHTLenS survey \citep{CFHTLenS-data}. We match our galaxies (see below) to the public CFHTLenS-W3 catalogue, within a 0.72~arcsec radius ($\approx 4$~pixels) on the sphere. We use their best-fit redshift measures \citep{2012MNRAS.421.2355H} (\texttt{Z\_B}) of the $576\,000$ matched galaxies to get an estimate for the redshift distribution $N(z)$ of our $r$-band catalogue. We then use this $N(z)$ for the entire galaxy sample. Given the observing strategy used for CFIS, which provides a constant depth over the observed area, this extrapolation seems justified. The distribution is shown in Fig.~\ref{fig:n_of_z}.
        
        This extrapolation introduces sample variance to the redshift distribution. \cite{2006APh....26...91V} estimated the variance that is added to the cosmic-shear covariance. In our case, this variance is sub-dominant compared to shape noise and cosmic variance, with the strongest contribution at scales of around $2\arcmin$.


\section{The \ShapePipe\ weak-lensing processing pipeline}
\label{sec:pipeline}

    ShapePipe is a modular weak-lensing processing and analysis pipeline written in Python. The package itself is comprised of two principal subpackages: one that constitutes the core pipeline architecture, and a second part that contains the various modules that account for each of the steps in the pipeline.
    
    ShapePipe makes use of the open-source package management system Conda\footnote{\url{https://docs.conda.io/}} to provide a version-controlled environment for each pipeline release. This ensures consistency in the versions of third-party packages employed by ShapePipe including those not written in Python ({\it e.g.} \textsc{SExtractor}, \textsc{PSFEx},  {\it etc.}). This in turn ensures a level of  reproducibility of ShapePipe results on different processing platforms.

    \subsection{Modular multiprocessing architecture}
    \label{sec:core}
    
        The core pipeline subpackage (hereafter referred to as the ShapePipe core) manages basic operations such as argument parsing, logging, dependency handling, and reading and writing FITS files. In addition to this, the pipeline core is responsible for parallel processing and module handling.
        
        The ShapePipe core operates under the assumption that a series of input products can be handled in an {\it embarrassingly parallel} manner via a series of independent jobs. The package uses two different methods for distributing jobs. On shared memory systems the ShapePipe core uses Joblib \citep{joblib20} to implement a form of symmetric multiprocessing (SMP), while on larger clusters a message passing interface (MPI) is implemented via MPI for Python \citep{dalcin05, dalcin08, dalcin11}. This enables ShapePipe to be run with as many CPU cores as are available on a given platform.
        
        The ShapePipe core was designed to be as modular as possible, meaning that various processing steps 
        could be updated, supplemented or replaced as new advances are made. A series of modules and the order in which they should be run can be specified in the configuration file before launching the pipeline. This enables ShapePipe to better adapt to data sets coming from different surveys.
        
        The ShapePipe core has been optimised to minimise the amount of memory used. This ensures that the majority of system resources are available for the modules.
        
        Finally, the ShapePipe core makes use of the following third-party packages to provide extra funcitonality: Astropy \citep{astropy:2018}, ModOpt \citep{farrens:20}, and Numpy \citep{numpy}.
    
    
        
        


\section{PSF modeling}
\label{sec:psf_section}
    
    The estimation of the point spread function (PSF) is a critical step for weak-lensing shape measurement. The PSF encompasses all aberrations induced in galaxy images due to optical imperfections and, most importantly, atmospheric effects such as refraction or turbulence. In the following section we describe how we model the PSF and its variations on the CFIS single-exposure images. The resulting model is used in Sect.~\ref{sec:shape_measurement} to correct for PSF effects in galaxy shapes.

    To model the PSF one needs to have a selection of stars, which should be as pure as possible and homogeneously distributed over the area of interest. Given the large dither and different observation periods between overlapping exposures, we decided to perform star selection at the single-exposure level. Despite their lower signal-to-noise ratio, the single exposures provide a more spatially stable distribution of stars. The star selection, as well as the PSF modeling, is performed independently on each of the 40 chips that constitute the \textsc{MegaCAM} CCD mosaic.

    \subsection{Star selection}
    \label{sec:star_selection}
    
        The star selection is a challenging step for UNIONS because, to date, only the $r$-band has been observed over an area and depth large enough to accurately model the PSF. Therefore, we select star candidates by identifying the stellar locus in a size-magnitude diagram, using the fact that the observed size of stars does not correlate with their luminosity.
            
        The selection is done in two steps. We first run \textsc{SExtractor}\footnote{\url{https://github.com/astromatic/sextractor}} \citep{1996A&AS..117..393B} with a relatively large area threshold to avoid polluting the star candidate sample with artefacts (bad pixels, cosmic ray hits, etc.). For that we fix the value \texttt{DETECT\_MINAREA = 10} \editt{pixels}. This corresponds to a circular aperture with a radius of $\approx 2$ pixels.
        
\begin{figure}
            \centering
            \includegraphics[width=1.\linewidth]{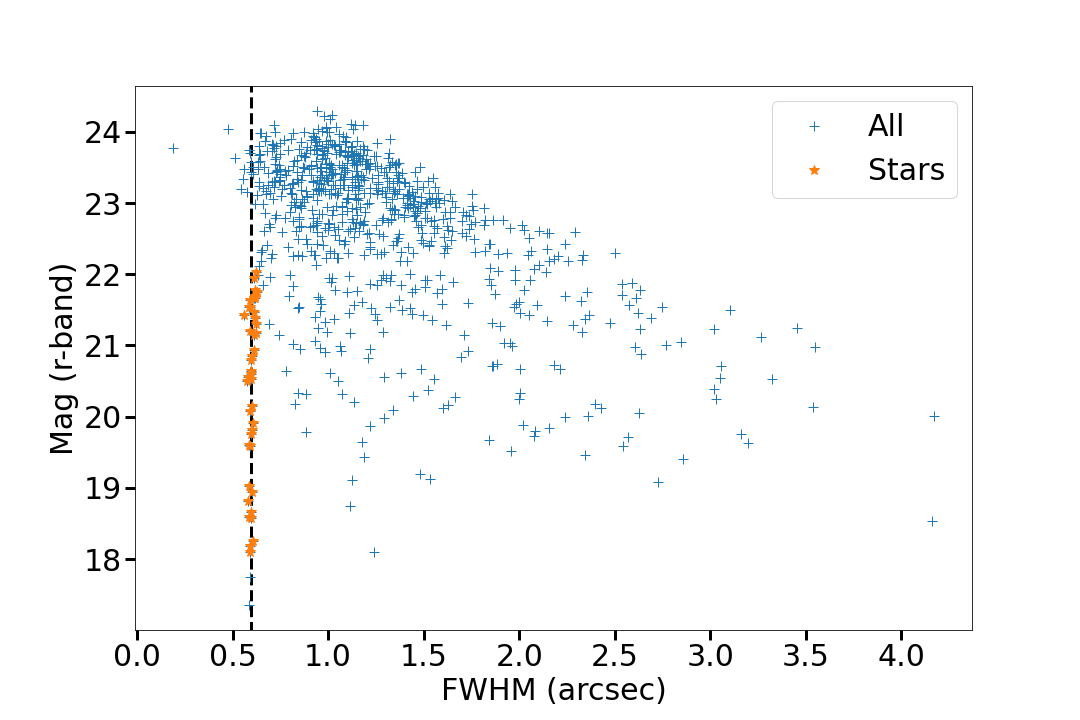}
            \caption{Size-magnitude diagram for one CCD. In \textit{orange} we show the selected stars. The \textit{dashed line} represents the mode of the stellar locus at $0.59$~arcsec, which is automatically estimated by \ShapePipe.}
            \label{fig:star_select}
        \end{figure}

\begin{figure}
            \centering
            \resizebox{\hsize}{!}{
                \includegraphics[width=1.\linewidth]{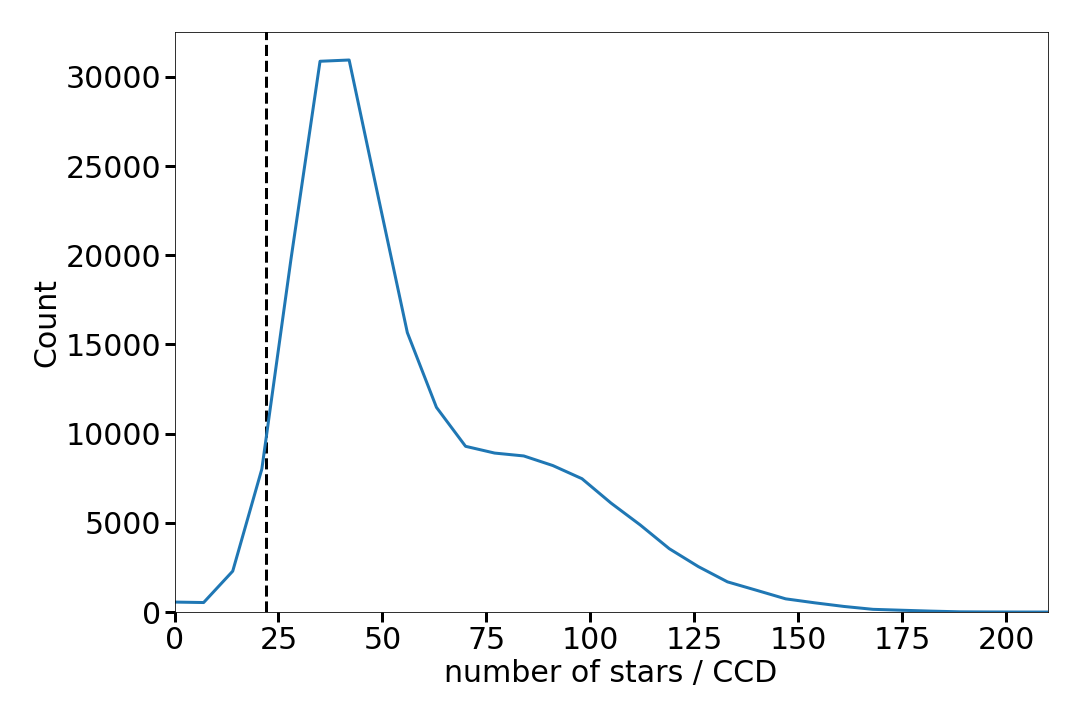}
            }
            \resizebox{\hsize}{!}{
                \includegraphics[width=1.\linewidth]{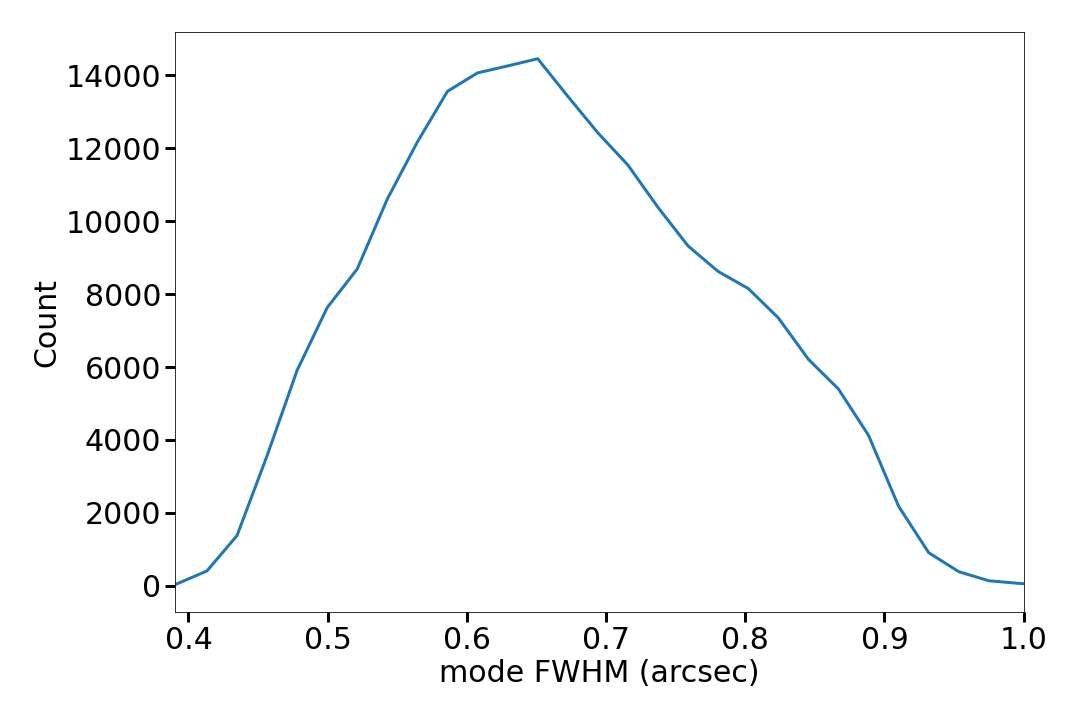}
            }
            \caption{Distribution of selected stars over all $208\,000$ CCDs. \textit{Top}: Number of stars per CCD. The dashed line represents the cut at 22 stars/CCD below which the CCD is discarded for the PSF estimation. \textit{Bottom}: Mode (per CCD exposure) of the FHWM of stars.}
            \label{fig:global_stats}
        \end{figure}

        Next, we select stars in a size-magnitude plane. We first pre-select a sample of objects with FWHM (full-width half maximum) between $0.3$ and $1.5$ arcsec, and compute the mode of the FWHM distribution, which provides us with an estimate of the stellar locus. From this pre-selection we keep objects for which the FWHM is within $0.04$ arcsec of the mode. In addition to these size cuts, we only use star candidates in the magnitude range $18 < r < 22$. These limits remove saturated stars, and faint, noisy objects that might be galaxies. This selection is carried out independently for each field of view, and also on each CCD. This accounts for the varying seeing between single exposures, and for the PSF size that changes with position on the focal plane. The star selection for one CCD is shown in Fig.~\ref{fig:star_select}. The stellar locus is clearly visible (in orange) and there are enough stars to reliably estimate the mode. The distribution of the number of selected stars per CCD is shown in the top panel of Fig.~\ref{fig:global_stats}.
        We set a threshold of 22 stars, below which a CCD will not be considered for PSF modelling, and consequently will not contribute to the multi-epoch shape measurements of galaxies imaged by this CCD. Only a small fraction of CCDs have a number of stars smaller than this threshold. The computation of the mode and star selection is performed automatically by the pipeline. \editt{\sout{The method is very robust, we have observed no failures for CCDs above the threshold.}}
        
        The bottom panel of Fig.~\ref{fig:global_stats} shows the FWHM distribution over all $208\,000$ CCDs, demonstrating the excellent image quality of CFHT/\textsc{MegaCAM}. With an average seeing of $0.65$ arcsec, UNIONS is at the same level as the HSC survey \citep{Mandelbaum_2017} and substantially better than what was achieved by DECam \citep{2017arXiv170801533Z}.
    
    \subsection{PSF estimation and interpolation}
        \label{sec:psf_estimation}
        %
        %
        
        In this section we describe how we model the PSF on each single exposure.
        We make use of the \textsc{PSFEx} software package\footnote{\url{https://github.com/astromatic/psfex}} \citep{PSFEx_paper}. Since we carry out our own star selection (see Sect.~\ref{sec:star_selection}), we disable the internal \textsc{PSFEx} pre-selection, and the PSF is thus obtained using the entire star sample. We ran \textsc{PSFEx} in \textsc{PIXEL} basis mode. This means that no assumptions are made regarding the parametric profile for the PSF. Rather, the pixels themselves are fitted in real space. This fit is initialised using the median profile of the sources provided in the field. The parameters fitted by \textsc{PSFEx} are the pixels that make up a set of PSF basis functions, $(S_{pq})_{\substack{p,q\ge0\\p+q\le d}}$, each of which is an image of the same size as the desired PSF model. These are then combined as a polynomial function of the position $(x,y)$ (here chosen to be the pixel position) to capture the spatial variations of the PSF:
        \begin{align}\label{eq:psfex}
            H(x,y) = \sum_{\substack{p,q\ge0\\p+q\le d}}x^py^qS_{pq}.
        \end{align}
        
        The maximal degree of the polynomials, $d$, is chosen through the PSFEx parameter \texttt{PSFVAR\_DEGREES}, which we set to two.
        
        \textsc{PSFEx} runs an iterative $\chi^{2}$ fit which allows for the removal of potential outliers at each iteration. This outlier rejection method removes around $0.08\%$ of all stars from the final PSF sample, demonstrating that our star selection (Sect.~\ref{sec:star_selection}) is robust.

        The PSFEx parameters we used are presented in Table \ref{table:psfex_param}. We have chosen not to over-sample the PSF models. This choice of parameters was driven by a comparison of PSF model performance using different sets of parameters and repeating the validation tests of Sect.~\ref{sec:psf_validation} on a large subset of the CFIS data.
        
        The computation of the \textsc{PSFEx} model at any given position (as in Eq.~\eqref{eq:psfex}) is normally carried out by \textsc{SExtractor}. However, since we perform our own shape measurement rather than relying on \textsc{SExtractor}, we use our own Python module to recombine the \textsc{PSFEx} basis functions.
        
        \begin{table}
			\centering
			\begin{tabular}{ |p{3cm}|p{4.3cm}|}
				\hline
				\multicolumn{1}{|l|}{Parameter} & \multicolumn{1}{|c|}{Value} \\
				\hline
				\hline
				\texttt{BASIS\_TYPE} & PIXEL \\
				\texttt{PSF\_SAMPLING} & 1 \\
				\texttt{CENTER\_KEYS} & XWIN\_IMAGE, YWIN\_IMAGE \\
				\texttt{PSFVAR\_KEYS} & XWIN\_IMAGE, YWIN\_IMAGE \\
				\texttt{PSFVAR\_DEGREES} & 2 \\
				\hline
			\end{tabular}
			\caption{\textsc{PSFEx} parametrisation. All other parameters are kept to their default values.}
				\label{table:psfex_param}
		\end{table}
		
\section{Shape measurement}
\label{sec:shape_measurement}

    In this section we first describe how the galaxy sample is selected using the spread model. Then, we present the method we used for measuring galaxy shapes. To calibrate the estimated shear, we make use of the metacalibration framework. The shape measurement is based on a joint multi-epoch\footnote{Uses several observation of the same object from different single exposures} model fitting and makes use of the \textsc{ngmix} software package\footnote{\url{https://github.com/esheldon/ngmix}} \citep{2017ApJ...841...24S}.
    
    \subsection{Source extraction}
    
        We start the processing by extracting all the sources using \textsc{SExtractor}, with the parametrization presented in Table \ref{table:sex_param}. We extract sources for which the pixel values are above $1.5$ times the noise variance. This is set with both parameters \texttt{THRESH\_TYPE} and \texttt{DETECT\_THRESH}. We do the extraction on stacked images which provide a better signal-to-noise ratio, and most artifacts have a reduced amplitude with respect to single exposures (due to time or position dependence). The detection is performed on a filtered image, for which we used the default 3x3 Gaussian kernel. This filtering smoothes the image and makes the detection less sensitive to noise fluctuation. As was done for the stars, we do not include sources that are too small (\texttt{DETECT\_MINAREA = 10}, see Sect.~\ref{sec:star_selection}). This choice could lead to a detection bias, but overall we are more confident that our sample contains astrophysical sources and not artefacts.
        
        \begin{table}
			\centering
			\begin{tabular}{ |p{3cm}|p{4.3cm}|}
				\hline
				\multicolumn{1}{|l|}{Parameter} & \multicolumn{1}{|c|}{Value} \\
				\hline
				\hline
				\texttt{THRESH\_TYPE} & RELATIVE \\
				\texttt{DETECT\_THRESH} & 1.5 \\
				\texttt{DETECT\_MINAREA} & 10 \\
				\texttt{FILTER} & Y \\
				\texttt{FILTER\_NAME} & kernel\_3x3.conv (default) \\
				\texttt{DEBLEND\_NTHRESH} & 32 \\
				\texttt{DEBLEND\_MINCONT} & 0.001 \\
				\hline
			\end{tabular}
			\caption{\textsc{SExtractor} parametrisation. All other parameters are kept to their default values.}
			\label{table:sex_param}
		\end{table}

    \subsection{Galaxy selection}
    \label{sec:galaxy_selection}
    
        Among the several techniques available to select galaxies we use the spread model introduced in \cite{Mohr_2012} and \cite{Desai_2012}. This method proposes to compare each extracted source to a point-source-like and an extended object. Sources with a spread model equal to zero are considered as point sources. A spread model larger than zero corresponds to extended objects, while values below zero corresponds to objects smaller than the PSF (\textit{i.e.} artefacts). As the spread model is noisy, we can construct an error for this measure, $\sigma_{s}$, which we use as a relaxation parameter to do the final classification. We do not consider the spread model to be sufficiently robust to classify PSF stars, which is why it is only used to pre-select our galaxy sample. Here we  encounter the same difficulty as for the star selection, in that we do not have access to colour information which would help to more accurately select our galaxy sample. However, as we demonstrate below, a conservative spread-model classification based on a single band is sufficiently accurate to yield a pure galaxy sample.
        Another difficulty at this stage is the handling of the PSF model. Indeed, as discussed in Sect.~\ref{sec:psf_estimation}, PSF models extracted from stacked images proved to be unreliable. Instead, we extrapolate the PSF information from the measurement on the single exposures to the stacked images. This implementation has been tested on a set of simulated images of stars and galaxies (which will be described in more detail in a future paper). The results show that we have only $0.7\%$ of miss-classified stars in our galaxy sample. 
        Here we present the equations of the spread model, $s$, and the spread model error, $\sigma_{s}$:
        \begin{equation}
            s = \frac{\vec{G}^{\mathrm{T}}~\mathrm{\vec{W}}~\vec{I}}{\vec{P}^{\mathrm{T}}~\mathrm{\vec{W}}~\vec{I}} -  \frac{\vec{G}^{\mathrm{T}}~\mathrm{\vec{W}}~\vec{P}}{\vec{I}^{\mathrm{T}}~\mathrm{\vec{W}}~\vec{P}};
            \label{eq:sm}
        \end{equation}
       %
        \begin{align}
            \sigma_{s} = &
            \frac{1}{(\vec{P}^{\mathrm{T}}~\mathrm{\vec{W}}~\vec{I})^{2}}
            \left( (\vec{G}^{\mathrm{T}} \mathrm{\vec{Cov}}~\vec{G})
            (\vec{P}^{\mathrm{T}}~\mathrm{\vec{W}}~\vec{I})^{2}(\vec{P}^{\mathrm{T}}~\mathrm{\vec{W}}~\vec{I})^{2}
            \right. \nonumber \\
            &
            + (\vec{P}^{\mathrm{T}}~\mathrm{\vec{Cov}}~\vec{P})(G^{\mathrm{T}}~\mathrm{\vec{W}}~\vec{I})^{2}
            \nonumber \\
            & \left.
            - 2(\vec{G}^{\mathrm{T}}~\mathrm{\vec{Cov}}~\vec{P})(\vec{G}^{\mathrm{T}}~\mathrm{W}~\vec{I})(\vec{P}^{\mathrm{T}}~\mathrm{\vec{W}}~\vec{I}) \right)^{1/2},
            \label{eq:sm_err}
        \end{align}
        with
        \begin{itemize}
            \item $\vec{P}$: PSF represented by an isotropic Gaussian with sigma equal to the mean sigma of the PSF model of the single epoch images interpolated to the position of the object detected on the stack.
            \item $\vec{G}$: Extended sources represented by an exponential profile with a scale radius of $1/16~\mathrm{PSF_{FWHM}}$ convolved by the PSF $\vec P$.
            \item $\vec{I}$: The image postage stamp of the object.
            \item $\vec{W}$: The weight image postage stamp.
            \item $\mathrm{\vec{Cov}}$: The covariance matrix of the noise, assumed to be diagonal,
            $\mathrm{\vec{Cov}}_{ij} = \delta_{ij} \vec{W}_{ii}^{-1}$
        \end{itemize}

\begin{figure}
            \centering
            \includegraphics[width=1.\linewidth]{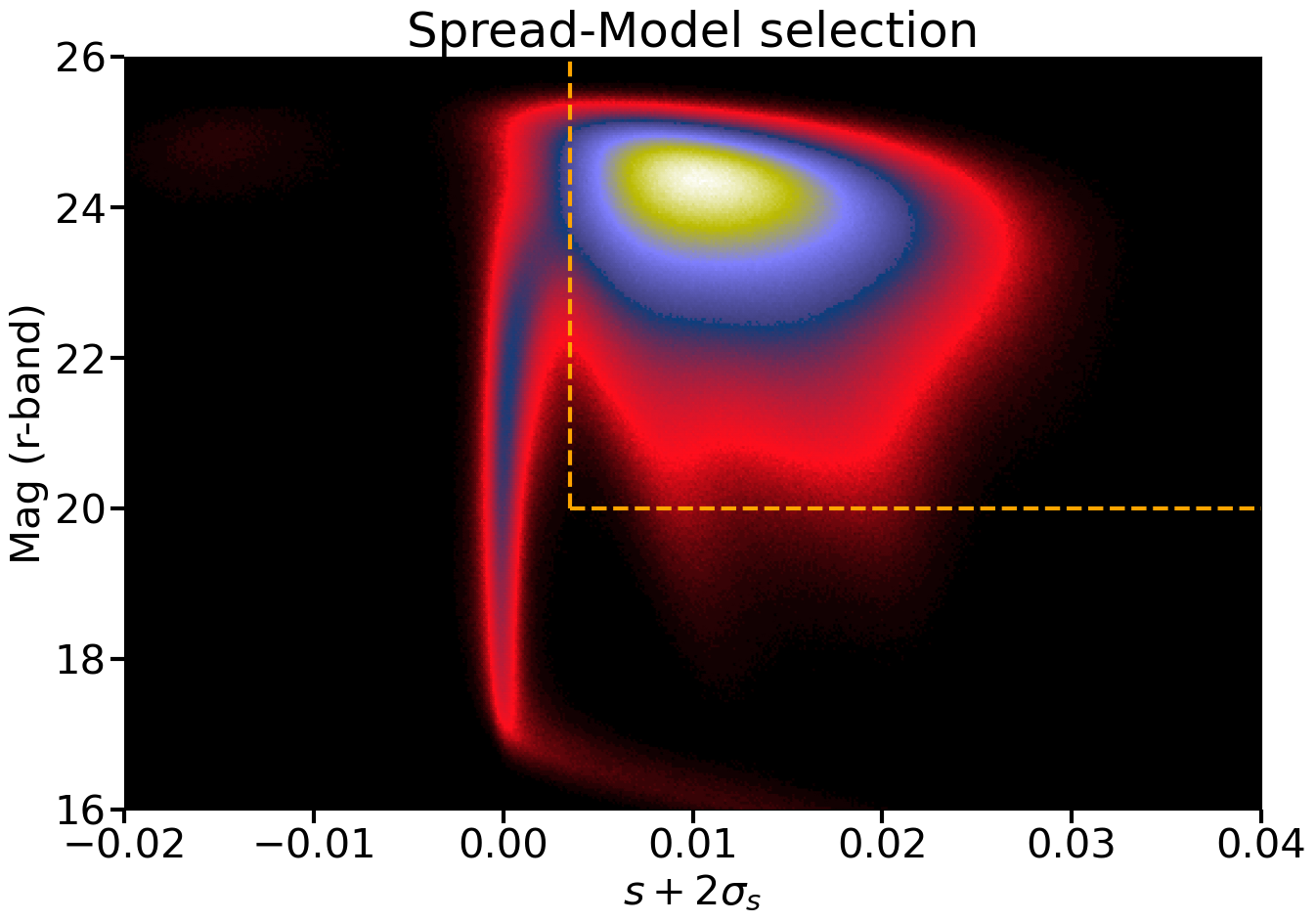}
            \caption{Spread model classification. The \textit{orange} area corresponds to the objects that have been selected for the galaxy sample. (The yellow contours represent a density 8 times larger than red).}
            \label{fig:sm_selection}
        \end{figure}

        We then use the spread model to make the selection presented below:
        \begin{itemize}
            \item $s + 2~\sigma_{s} > 0.0003$
            \item $s > 0$
            \item $20 < \mathrm{MAG\_AUTO} < 26$,
        \end{itemize}
        where $s$ is given by Eq.~\eqref{eq:sm}, and $\sigma_{s}$ by Eq.~\eqref{eq:sm_err}.
        The cuts are illustrated in Fig.~\ref{fig:sm_selection}.
        
        In addition to the spread model, we also apply cuts based on quantities derived from the model fitting described in Sect.~\ref{sec:me_shapes}. The spread model gives a sample of extended sources, which needs to be refined. We apply additional cuts to the ratio between the size of the galaxies, $T_{\mathrm{gal}}$, and the size of the PSF, $T_{\mathrm{PSF}}$. We also remove objects that have a low SNR, which we defined as the ratio between the measured flux of the galaxies, $F$, and the error on the flux, $\sigma_{F}$. The cuts are:
        \begin{itemize}
            \item $\frac{T_{\mathrm{gal}}}{T_{\mathrm{PSF}}} > 0.5$
            \item $10 < \frac{F}{\sigma_{F}} < 500$
        \end{itemize}

    \subsection{Metacalibration}
    
        Metacalibration is a method introduced by \cite{2017arXiv170202600H} and was used in the DES collaboration \citep{2017arXiv170801533Z}. This method allows one to calibrate shear measurements without the need to create a large number of time-consuming image simulations. It consists in measuring the response, $\mat R$, of a shape measurement algorithm to a shear artificially applied to an image. To introduce this response we start with the classic equation in the weak-lensing limit used to estimate the mean shear, $\gamma$, from an ensemble of galaxies. With $i \in [1,2]$ denoting the ellipticity component, the observed ellipticity, $e_{i}^{\mathrm{obs}}$, is given as:
        \begin{equation}
        \label{eq:def_e_obs}
            \left\langle e^{\mathrm{obs}}_{i} \right\rangle = (1 + m_i) \left\langle \gamma_{i} \right\rangle + c_i,
        \end{equation}
        with $m_{i}$ and $c_{i}$ being the multiplicative and additive bias for component $i$. This relation assumes that the intrinsic ellipticity of galaxies vanishes on average.
        
        The shear response generalises this equation, and relates the observed ellipticity to shear for an individual galaxy of intrinsic ellipticity $e^{\mathrm{int}}_{i}$, as:
        \begin{equation}
            e^{\mathrm{obs}}_{i} = e^{\mathrm{int}}_{i} + \sum^{2}_{j=1} \mat{R}_{ij} \gamma_j + c_{i},
        \end{equation}
        where $\mat R$ is the response matrix, which can be described as:
        \begin{equation}
            \mat{R}_{ij} = \frac{\partial e_{i}^{\mathrm{obs}}}{\partial \gamma_{j}},
        \end{equation}
        and approximated to:
        \begin{equation}
            \mat{R}_{ij} = \frac{e_{i}^{+} - e_{i}^{-}}{2 \Delta g_{j}},
        \end{equation}
        where $e_{i}^{\pm}$ represents the $i^{\mathrm{th}}$ component of ellipticity measured on an image with added shear component $\pm \Delta g_i$.
        
        To apply an artificial shear, the image needs to be first deconvolved by its original PSF. The deconvolution is performed by a division in Fourier space. We can then apply an artificial shear and reconvolve the object by a larger PSF\footnote{Using a larger PSF for the reconvolution ensures that the artifacts created by the deconvolution due to the presence of noise in the images are smoothed out.}. Finally, we generate a noise image (\textit{i.e.} a postage stamp of the same size as the original observation but containing only noise) with the same variance as the original image to cancel the correlations created by the shearing process. To cancel the correlations, the noise image undergoes the same process of deconvolution, shearing and reconvolution. The difference being in the shearing. To cancel the correlations created in the image we apply the shear on a $90\deg$ rotated version of the noise image, which is then rotated back, and applied to our science image (for a more in-depth description of the noise handling we refer the reader to Sect. 4.2 of \cite{2017ApJ...841...24S}). In this work, we use the \textsc{ngmix} software package to handle the Metacalibration steps.
        
        This method creates four images used for the calibration, and one for the measurement. The response is composed of two components, $\mat R = \langle \mat R^{\mathrm{shear}} \rangle +  \langle \mat R^{\mathrm{selection}} \rangle$. $\langle \mat R^{\mathrm{shear}} \rangle$ represents the corrections of the shear bias, which here encompass model bias and noise bias. $\langle \mat R^{\mathrm{selection}} \rangle$ accounts for the biases due to selection cuts (\textit{e.g.} on magnitude or object size). To correct for these selection effects, the cuts have to be performed on parameters obtained from the sheared images. We refer to these cuts as the selection mask, $M^\pm$. Both, $\langle \mat R^{\mathrm{shear}} \rangle$ and $\langle \mat R^{\mathrm{selection}} \rangle$ can be defined as follows:
        \begin{equation}
            \langle \mat{R}_{ij}^{\mathrm{shear}} \rangle = \left\langle \frac{e_{i}^{+} - e_{i}^{-}}{2 \Delta g_{j}} \right\rangle;
        \end{equation}
        \begin{equation}
            \left\langle \mat{R}_{ij}^ {\mathrm{selection}} \right\rangle = \frac{\langle e_{i}^{0,M^{+}} \rangle - \langle e_{i}^{0,M^{-}}\rangle }{2 \Delta g_{j}},
            \label{eq:R_selection}
        \end{equation}
        where $\langle e_{i}^{0,M^{\pm}} \rangle$ represents the average ellipticity measured on the image with no shear applied but using the selection mask (described in Sect.~\ref{sec:galaxy_selection}), $M^{\pm}$, from the images with a small shear $\pm \Delta g_i$ applied. Since the quantities to which we apply cuts are correlated to the shear, we will obtain different values for $\langle e_{i}^{0,M^{\pm}} \rangle$ depending on which sheared version of the image we apply the cuts. This leads to a non-zero selection response. Finally, we have $\Delta g_{j} = |\pm g_j|$, with $g_j = 0.01$ \editt{\sout{in our case}, according to \cite{Sheldon_2020}}.

        Here we measure the selection effects due to the different cuts one can apply to the shape catalogue. The cuts use the same criteria for all the sheared versions but they might give a different selection for each sheared version of the objects. The differences account for this effect, which is the selection effect we want to capture. It is important to note that only effects due to cuts on quantities measured on the sheared images can be accounted for. For example, it is not possible to correct for detection effects at this stage \citep{Sheldon_2020}.

    \subsection{Multi-epoch model fitting shape measurement}
    \label{sec:me_shapes}
    
        To measure the shapes of objects we use \textsc{ngmix}, a model-fitting technique. This method consists in finding the best-fit parameters for a model that minimises the function:
        \begin{equation}
            \chi^{2} = \sum_{i=1}^{n_{\textrm{epoch}}} ~\sum_{j=0}^{n_{\textrm{pixel}}} (I_{i,j} - (G_{j}*P_{i,j}))^{2} W_{i,j},
            \label{eq:chi2_shape_measurement}
        \end{equation}
        where $I_{i,j}$ corresponds to the vectorized image and $W_{i,j}$ to the weight describing noise variations in the image, for the single exposure $i$ and pixel index $j$. $G_{j}$ is the vectorized modelled image convolved ($*$) with the PSF of the corresponding single exposure $i$, $P_{i, j}$. Here, we model the  galaxies with a simple Gaussian profile. Despite being very simple, the model bias \citep{model_bias} is small. Furthermore, since this bias is calibrated to a large extent by the metacalibration framework. We quantify this in Sect.~\ref{sec:simulation}. The minimization (Eq.~\eqref{eq:chi2_shape_measurement}) is performed in Fourier space for computational reasons.

        The model is created using \textsc{GalSim}\footnote{\url{https://github.com/GalSim-developers/GalSim}} \citep{2015A&C....10..121R}. Six parameters are used to describe the model:
        \begin{itemize}
            \item Centroid shifts in both directions $\Delta x, \Delta y$;
            \item Two ellipticity components $e_{1}, e_{2}$;
            \item Half-light radius $r_{50}$;
            \item Flux $F$.
        \end{itemize}
        The model takes into account the optical distortions in all single exposures through the WCS (World Coordinates System) framework.
        
        For the minimization of $\chi^{2}$, a least-square algorithm is used with the following constraints:
        \begin{itemize}
            \item $r_{50} > 0.0001$ arcsec,
            \item $\sqrt{e_{1}^{2} + e_{2}^{2}} \leq 1$,
        \end{itemize}
       and the following priors:
       \begin{itemize}
           \item $e_{1,2}$: distribution from \cite{Bernstein_2014},
           \item $\Delta x, \Delta y$: Gaussian distribution centred on 0 with $\sigma = \mathrm{pixel\ scale} \approx 0.187$ arcsec,
           \item $r_{50}$: flat distribution in $[-10, 10^{6}]$ arcsec,
           \item $F$: flat distribution in $[-10^{4}, 10^{9}]$.
       \end{itemize}
        Finally, in order to reach convergence during the fitting operation we have to provide accurate initial values for the size and the flux. We thus run first an adaptive moments algorithm to initialise the least-square operation. In order to avoid any bias due to the choice of the initialisation, this is performed on each sheared version of the object. The initial values are chosen as follows:
        \begin{itemize}
            \item $\Delta x = \frac{Q_{1,0}}{Q_{0,0}}, \quad \Delta y = \frac{Q_{0,1}}{Q_{0,0}}$,
            \item $e_{1,2} = 0$,
            \item $r_{50} = \sqrt{2 \, \mathrm{ln} 2}  \sqrt{\frac{Q_{0,2} + Q_{2,0}}{2}}$,
            \item $F = Q_{0,0}$,
        \end{itemize}
        where
        \begin{equation}
            Q_{i,j} = \int (x - x_{0})^{i} I(x) w(x) (x - x_{0})^{j}{\rm d}^{2}x,
        \end{equation}
        represent the moments of the light profile $I(x)$ of order $i+j$ and weighted by a Gaussian window $w(x)$. The moments are computed using the HSM algorithm \citep{galsim_hsm} from the \textsc{GalSim} software package.
        
        We need to apply weights to the ellipticity values in order to take into account the measurement uncertainties. These weights are computed as:
        \begin{equation}
        \label{eq:ell_weight}
            w = \frac{1}{2\sigma_{\mathrm{SN}}^{2} + \sigma_{e_{1}}^{2} + \sigma_{e_{2}}^{2}},
        \end{equation}
        where $\sigma_{\mathrm{SN}}$ represents the raw shape noise, which has been measured from data to be 0.34 (here we used a simple variance estimator, a more precise estimate is given by Eq.~\eqref{eq:shape_noise}). $\sigma_{e_{1,2}}$ is the variance of the ellipticity parameters estimated during the model fitting.
        
\section{Diagnostics}
\label{sec:diagnostics}

    Here we present the validation tests performed on the PSF model and on the shape measurements. These tests are crucial to demonstrate a low enough level of  systematics for reliable science applications of the weak-lensing data.
    
    First, we present two sets of tests of the PSF model, some qualitative and some quantitative. For the shape measurement we have focused on the PSF residuals propagated to the galaxy shapes. Since no cosmological requirements are defined yet in the UNIONS collaboration, we have taken other surveys as a reference.
    
    \subsection{Ellipticity correlation functions}
    \label{sec:2pcf}
    
        Some diagnostics of the impact of the PSF on shape measurements concern the correlation between the measured ellipticities of different galaxy and star samples. First,
        we define the two-point correlation function $\rho$ of the ellipticity $e$, written as a complex number $e = e_1 + \mathrm{i} e_2$, of two samples. We consider two samples $A$ and $B$. If their ellipticities, as random fields on the sky at positions $\vec \theta$, $e^A(\bm \theta)$ and $e^B(\bm \theta)$, respectively, are statistically homogeneous and isotropic, their two-point function only depends on a scalar distance $\theta$, $\rho = \rho(\theta) = \langle {e^A}^\ast e^B \rangle(\theta)$. Here, the asterisk `*' denotes complex conjugation. Replacing the ensemble average by a spatial average over positions $\bm \theta^\prime$, we can write $\rho(\theta) =  \langle {e^A}^\ast(\bm \theta^\prime) e^B(\bm \theta^\prime + \bm \theta) \rangle$.
        An unbiased estimator of 
        $\rho$ is
        \begin{equation}
                \hat\rho(\theta) = \frac{\sum_{ij} w_i w_j {e^\ast}^A_i e^B_j }{\sum_{ij} w_i w_j} .
                \label{eq:rhopm}
        \end{equation}
        The weighted sum is carried out over pairs $(ij)$ of objects at sky positions $\bm \theta_i$, $\bm \theta_j$, whose distance $\theta_{ij} = |\bm \theta_i - \bm \theta_j|$ 
        is close to $\theta$. We use logarithmic bins, and therefore the angular bin is given by $\ln \theta - \Delta \ln \theta/2 < \ln \theta_{ij} < \ln \theta + \Delta \ln \theta/2$. The weights $w_i$ are defined in Eq.~\eqref{eq:ell_weight}.
        
        In the case of cosmic-shear, where the ellipticity samples $A$ and $B$ are galaxy shear estimates, the correlation function $\rho$ defined above is represented by $\xi_+$.
        In addition, $\xi_-$ is defined as $\xi_-(\theta) = \Re \left[ \langle {e^A} (\bm \theta^\prime) e^B(\bm \theta^\prime + \bm \theta) \exp(-4 \mathrm{i} \varphi ) \rangle \right]$, where $\varphi$ is the polar angle of $\bm \theta_{ij}$ \citep{SvWKM02}. We will make use of both functions in Sects.~\ref{sec:PSF_leakage} and \ref{sec:cosebis}.

    \subsection{PSF validation tests}
    \label{sec:psf_validation}
    
        In this section we present the tests we have performed on our PSF model to quantify the systematics. To be able to perform these tests properly, our star sample has been randomly divided in two:
        \begin{itemize}
          \item 80\% \editt{\sout{for the model estimation.} for creating the model.}
          \item 20\% for testing.
        \end{itemize}
        For the following validation tests we only use the test sample. \editt{In doing so, we can test the interpolation of the PSF model and at the same time be less sensitive to over-fitting.}
        
        \subsubsection{Focal-plane PSF residuals}
        \label{sec:psf_residual_test}
        
\begin{figure*}
                \centering
                \begin{minipage}{0.475\textwidth}
                    \centering
                    \includegraphics[width=1.\linewidth]{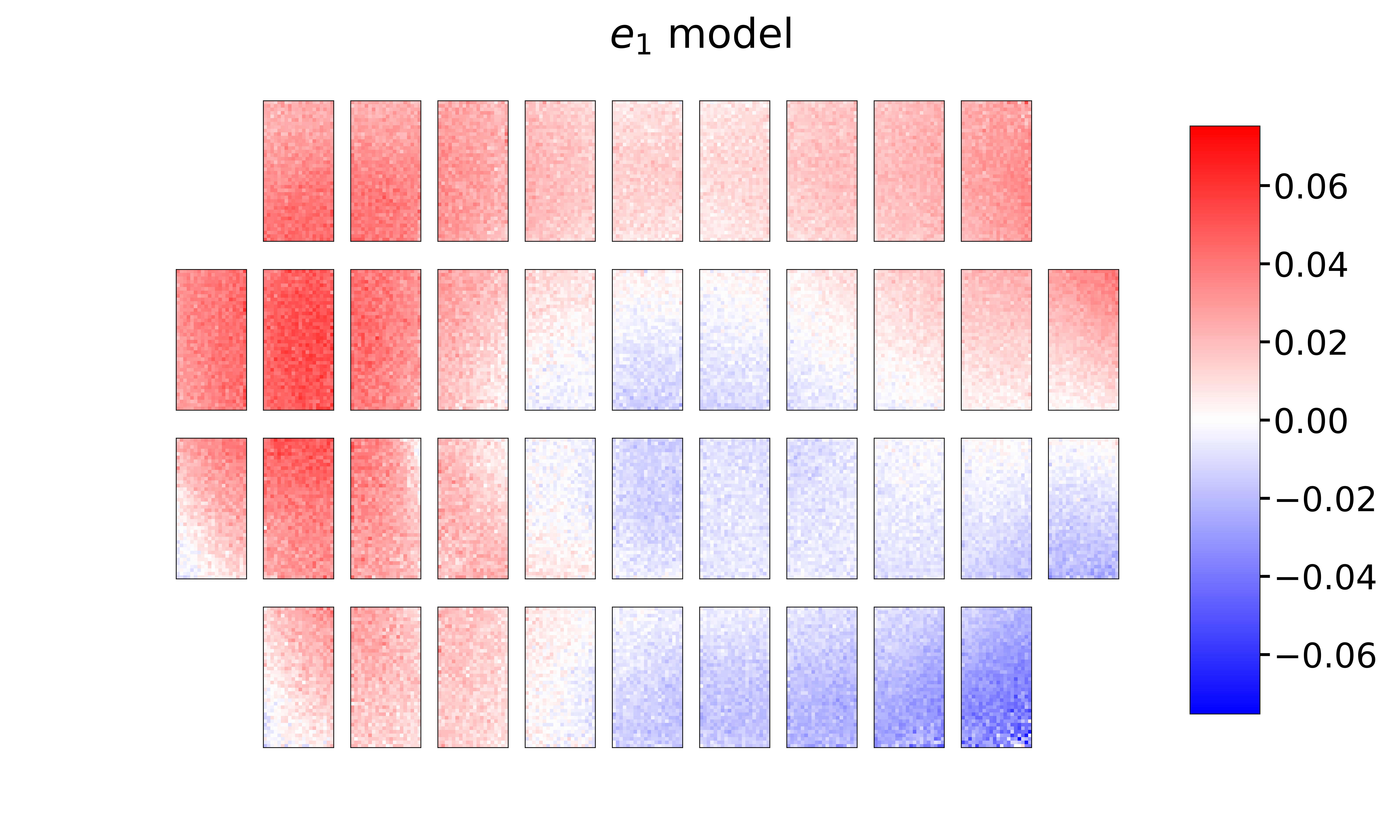}
                \end{minipage}
                \hfill
                \begin{minipage}{0.475\textwidth}
                    \centering 
                    \includegraphics[width=1.\linewidth]{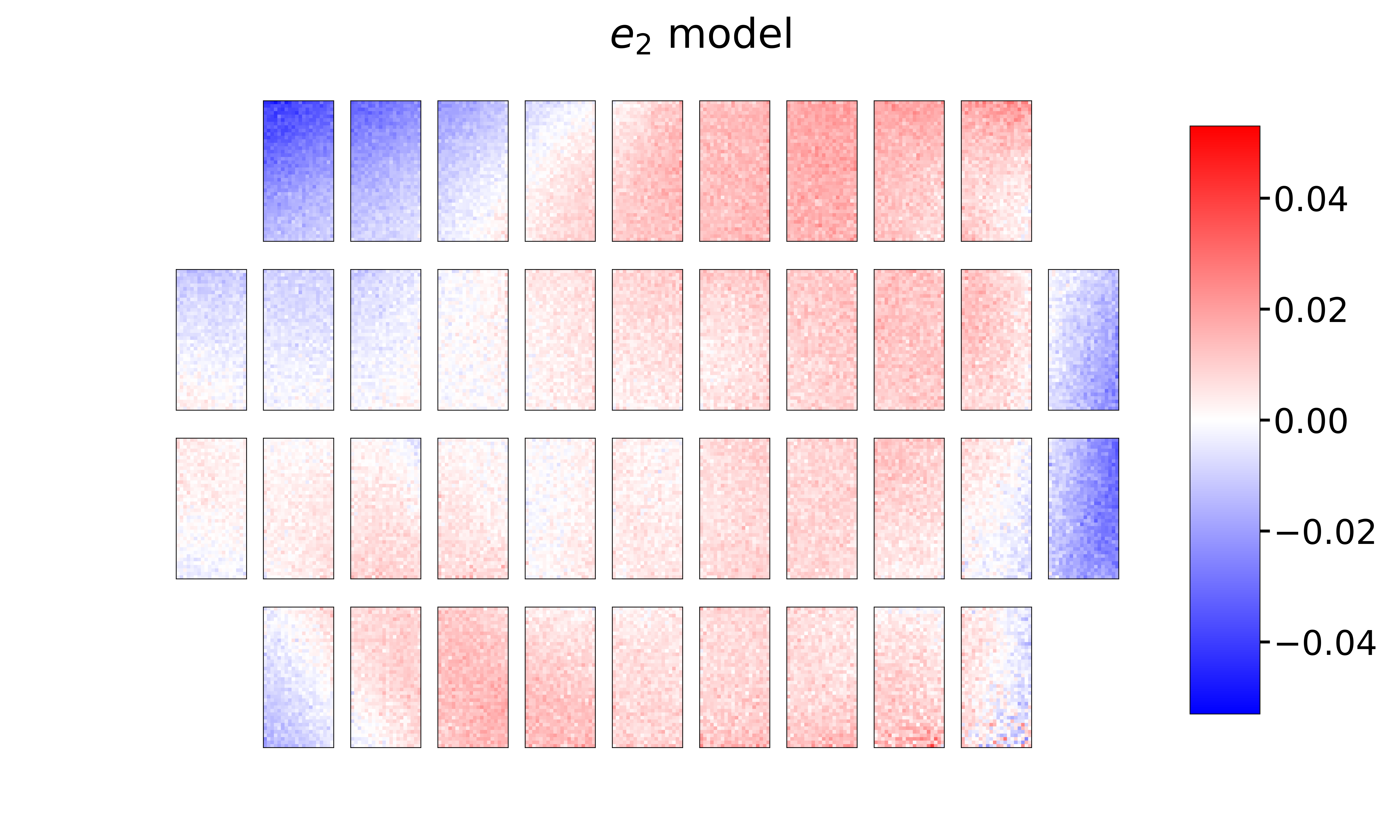}
                \end{minipage}
                \vskip\baselineskip
                \begin{minipage}{0.475\textwidth}
                    \centering 
                    \includegraphics[width=1.\linewidth]{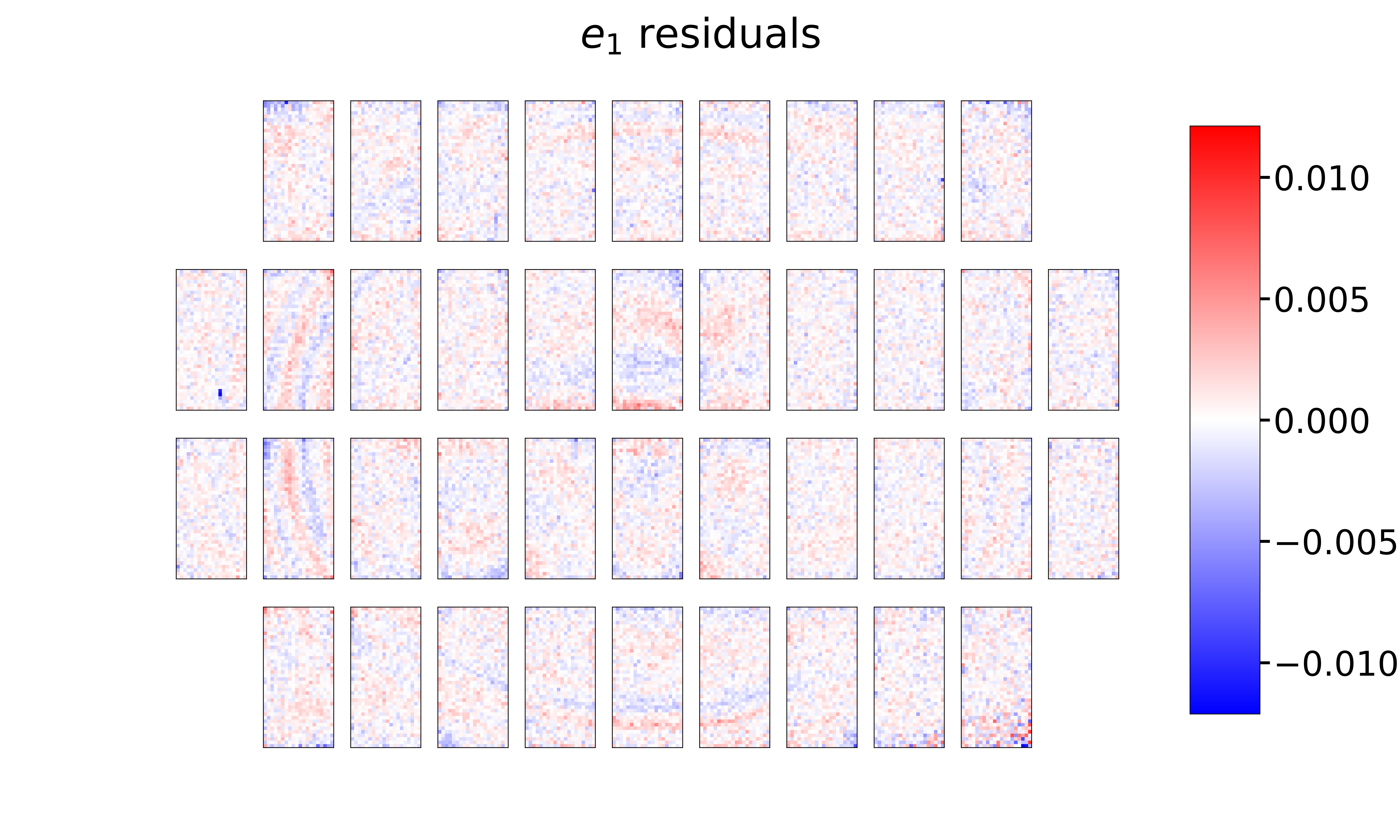}
                \end{minipage}
                \quad
                \begin{minipage}{0.475\textwidth}
                    \centering 
                    \includegraphics[width=1.\linewidth]{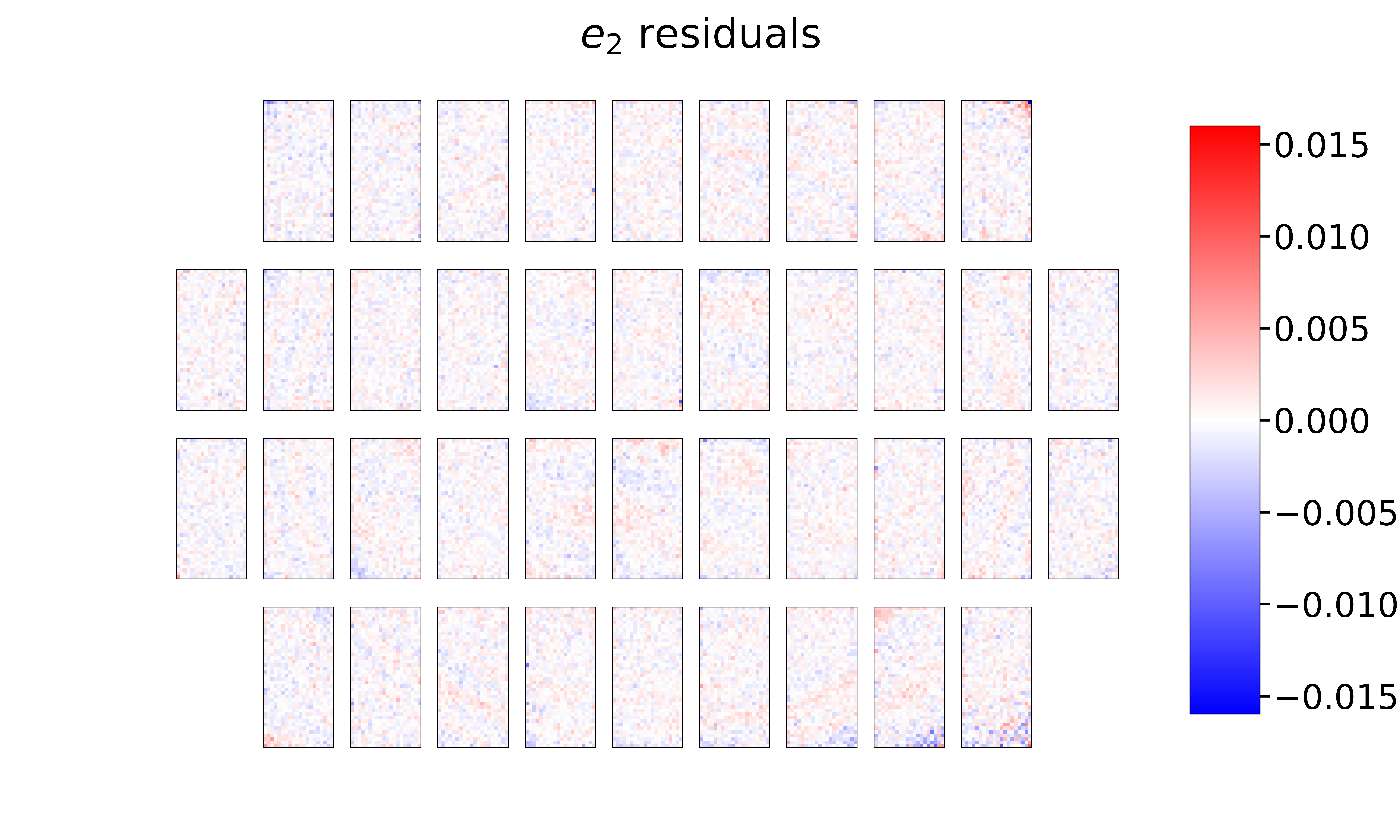}
                \end{minipage}
                \caption{PSF ellipticity (\textit{upper panels}), $e^{\mathrm{PSF}}$, and PSF ellipticity residuals (\textit{lower panels}), $\delta e^{\mathrm{PSF}}$, maps of the 40 CCDs of the \textsc{MegaCAM} focal plane.} 
                \label{fig:psf_res}
            \end{figure*}

\begin{figure*}
                \centering
                \begin{minipage}{0.475\textwidth}
                    \centering
                    \includegraphics[width=1.\linewidth]{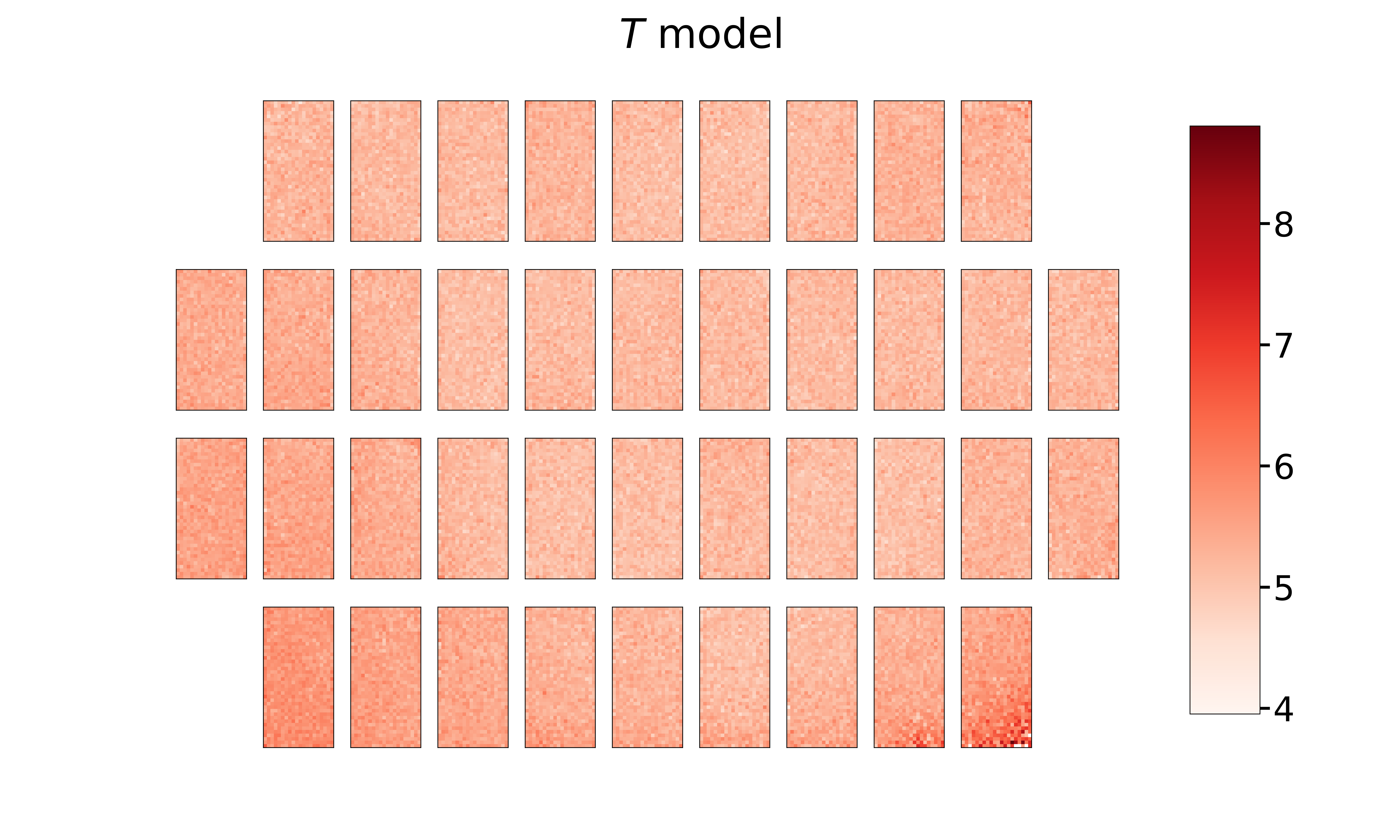}
                \end{minipage}
                \hfill
                \begin{minipage}{0.475\textwidth}
                    \centering 
                    \includegraphics[width=1.\linewidth]{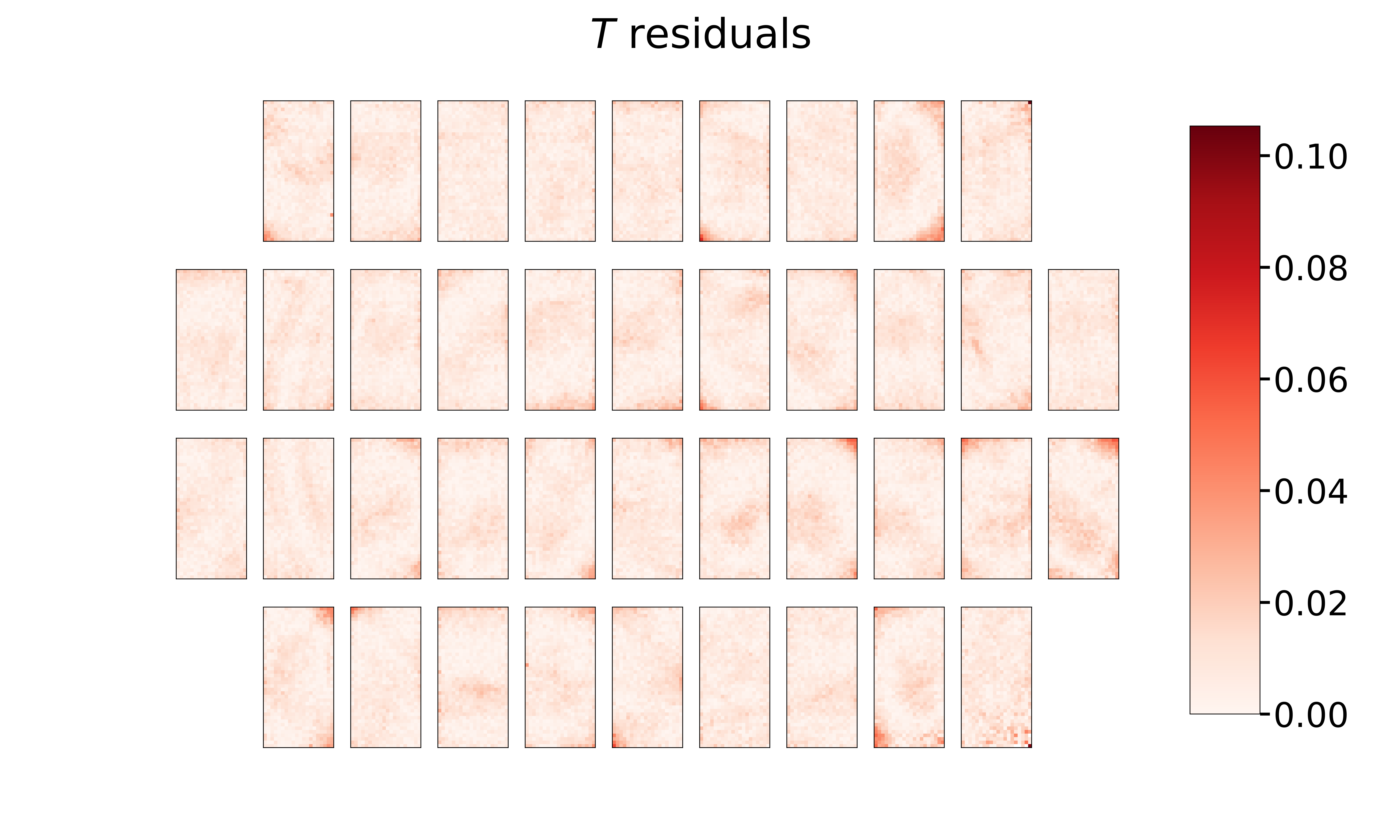}
                \end{minipage}
                \caption{PSF size (\textit{left panel}), $T_{\mathrm{PSF}}$, and PSF size residuals (in pixels$^2$) (\textit{right panel}), $\delta T_{\mathrm{PSF}}/T_{\mathrm{star}}$, maps of the 40 CCDs of the \textsc{MegaCAM} focal plane.} 
                \label{fig:psf_res_size}
            \end{figure*}

            To estimate the errors due to the PSF modelling, we first look at the ellipticity/size residuals between the model and the stars. In this work we define the residuals as $\delta e_{\mathrm{PSF}} = e_{\mathrm{PSF}} - e_{\mathrm{star}}$ with $e_{\mathrm{PSF}}$ and $e_{\mathrm{star}}$ being the ellipticity of the PSF at the star positions and the ellipticity of stars, respectively, and $\delta T_{\mathrm{PSF}} = T_{\mathrm{PSF}} - T_{\mathrm{star}}$ where $T_{\mathrm{PSF}}$ and $T_{\mathrm{star}}$ are the size of the PSF and the size of the star, respectively. The size is defined as follows:
            \begin{equation}
                T_{\mathrm{PSF/star}} = 2\sigma_{\mathrm{PSF/star}}^{2}.
            \end{equation}
            The ellipticity and the size ($\sigma_{\mathrm{PSF/star}}$) here are measured using the adaptive moments implementation from the HSM module in the \textsc{GalSim} software package. The model ellipticity and residuals are plotted in Fig.~\ref{fig:psf_res} while Fig.~\ref{fig:psf_res_size} presents the same statistics for the size. Each panel shows the average over all exposures as a function of position on the focal plane. For each of the 40 \textsc{MegaCAM} CCDs, values are averaged over pixels of size $20~\mathrm{arcsec}^2$.
            
            The PSF ellipticity residuals typically are a factor $10$ smaller than the PSF amplitude, which indicates an accurate PSF model on average. The PSF size is more uniform and reproduced by the model to a  higher accuracy, with the highest residual being situated at the CCD edges. 
            
            In the figure large circular patterns in the residuals can be seen. This could be caused by the low degree ($2$ in our case) of the polynomial used to construct the model. In this case, the patterns would reflect higher-order spatial variations. We have, however, tried to increase the degree of the polynomial and, due to the small number of selected stars, the resulting model was too noisy to be used. Another possibility is that this is caused by the PSF model being limited to each inidividual CCD. Since the patterns are larger than the size of one CCD, they might not be captured by this model. In upcoming work we will use the multi-CCD method presented in \citet{2020arXiv201109835L}, which produces smaller residual errors with respect to \textsc{PSFEx}.
            
        \subsubsection{$\rho$-statistics}
        \label{sec:psf_rho_stats}
        
            Another test one can do is to use the metrics proposed by \citet{2010MNRAS.404..350R} and \cite{2016MNRAS.460.2245J}. This consists in computing the spatial correlations between the ellipticities of the PSF model and residuals. For that purpose, the two-point correlation functions defined in Sect.~\ref{sec:2pcf} are used. The different $\rho$-statistics correlation functions are given as follows:
            \begin{equation}
                \rho_{1}(\theta) = \langle \delta e_{\mathrm{PSF}}^{*}(\bm{\theta}^\prime) \delta e_{\mathrm{PSF}}(\bm{\theta}^\prime + \bm{\theta}) \rangle;
            \end{equation}
            \begin{equation}
                \rho_{2}(\theta) = \langle e_{\mathrm{PSF}}^{*}(\bm{\theta}^\prime) \delta e_{\mathrm{PSF}}(\bm{\theta}^\prime + \bm{\theta}) \rangle;
            \end{equation}
            \begin{equation}
                \rho_{3}(\theta) = \left \langle \left(e_{\mathrm{PSF}}^{*} \frac{\delta T_{\mathrm{PSF}}}{T_{\mathrm{PSF}}}\right)(\bm{\theta}^\prime) \left(e_{\mathrm{PSF}} \frac{\delta T_{\mathrm{PSF}}}{T_{\mathrm{PSF}}}\right)(\bm{\theta}^\prime + \bm{\theta}) \right \rangle;
            \end{equation}
            \begin{equation}
                \rho_{4}(\theta) = \left \langle \delta e_{\mathrm{PSF}}^{*}(\bm{\theta}^\prime) \left(e_{\mathrm{PSF}} \frac{\delta T_{\mathrm{PSF}}}{T_{\mathrm{PSF}}}\right)(\bm{\theta}^\prime + \bm{\theta}) \right \rangle;
            \end{equation}
            \begin{equation}
                \rho_{5}(\theta) = \left \langle e_{\mathrm{PSF}}^{*}(\bm{\theta}^\prime) \left(e_{\mathrm{PSF}} \frac{\delta T_{\mathrm{PSF}}}{T_{\mathrm{PSF}}}\right)(\bm{\theta}^\prime + \bm{\theta}) \right \rangle.
            \end{equation}
            Note that some of these correlation functions use ellipticities weighted by the relative size residuals, $\delta T_{\mathrm{PSF}} / T_{\mathrm{PSF}}$.
            
\begin{figure*}
            \centering
            \resizebox{\hsize}{!}
            {
                \begin{minipage}{.5\textwidth}
                  \centering
                  \includegraphics[width=1\linewidth]{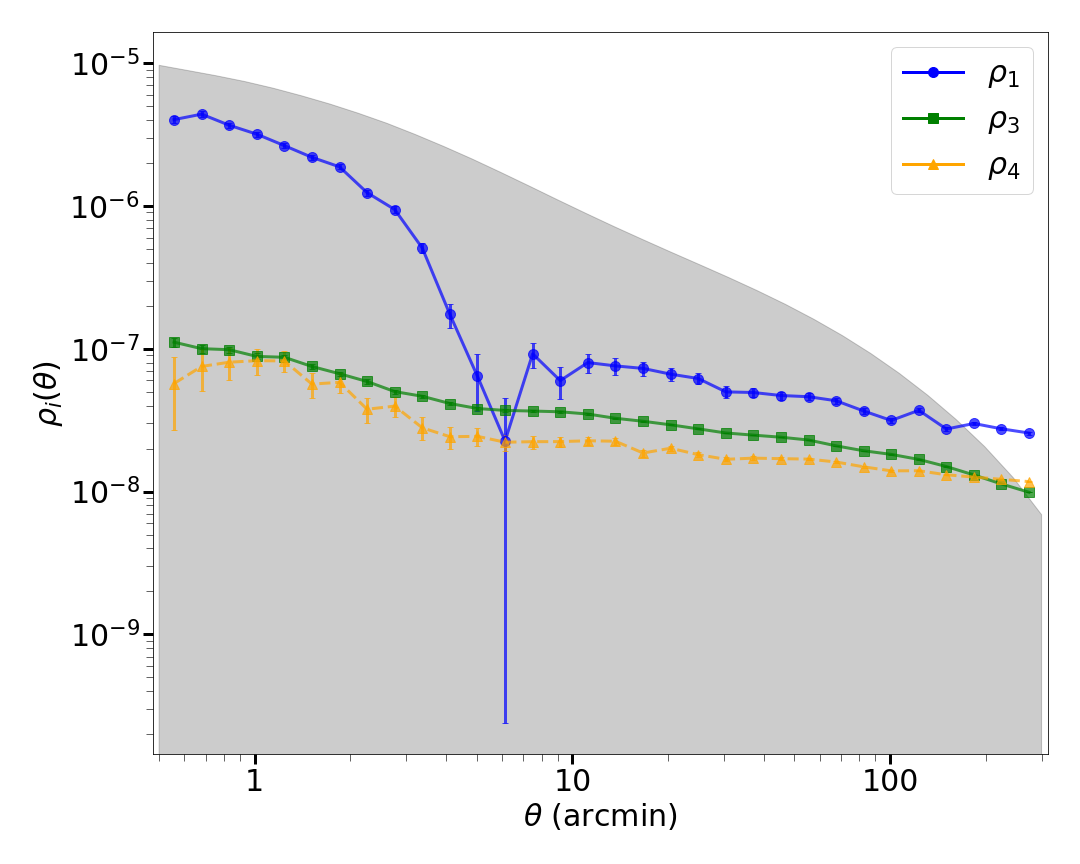}
                \end{minipage}
                \begin{minipage}{.5\textwidth}
                  \centering
                  \includegraphics[width=1\linewidth]{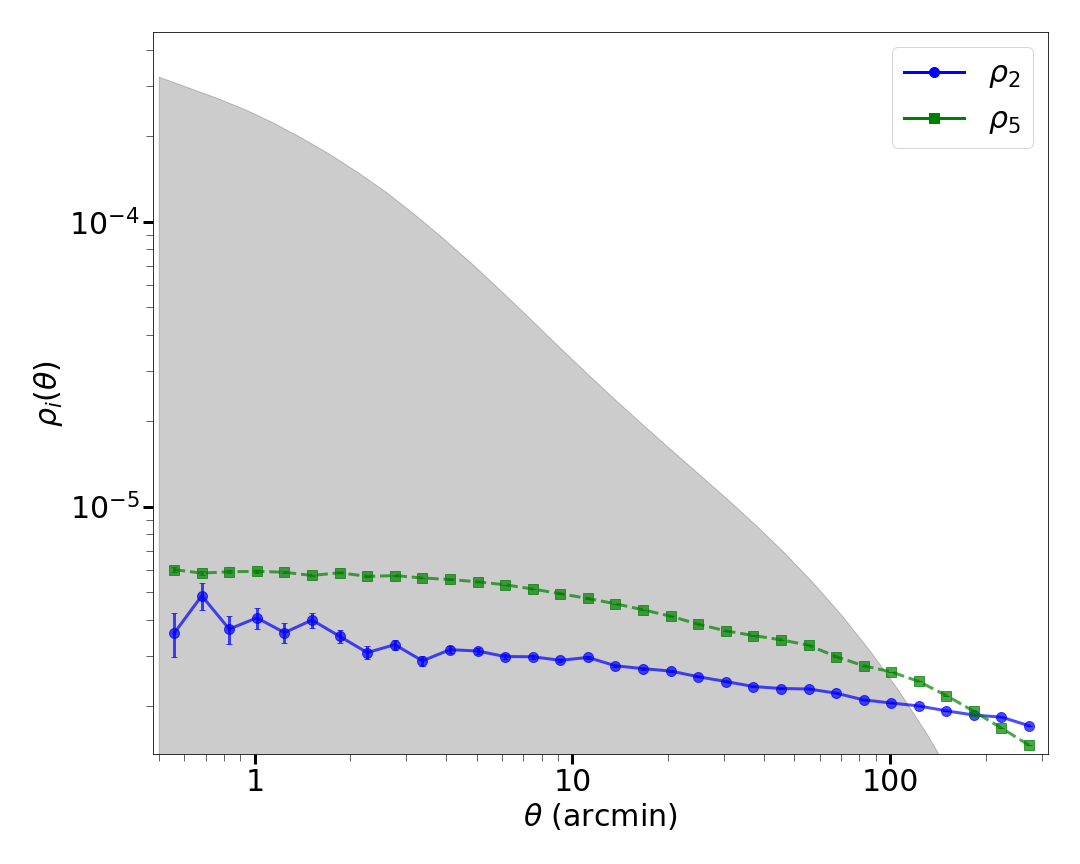}
                \end{minipage}
            }
            \caption{%
            The PSF residuals $\rho$-statistics. The \textit{grey} area are the requirement described in Sect.~\ref{sec:psf_residual_test}. Here we set $T_{\mathrm{PSF}}/T_{\mathrm{gal}} = 1$ and $\alpha = 0.1$.}
            \label{fig:rho_stat}
            \end{figure*}

            The $\rho$-statistics can be related to cosmology. They add directly to the measured shear two-point correlation function $\xi_+$. To constrain cosmological parameters from $\xi_+$ at a given precision, the amplitude of the $\rho$-statistics should not exceed an amount that can be computed as follows (\cite{2016MNRAS.460.2245J}):
            \begin{equation}
                |\rho_{1,2,3}(\theta)| < \left \langle \frac{T_{\mathrm{PSF}}}{T_{\mathrm{gal}}} \right \rangle^{-2} \delta \xi_{+}^{\mathrm{max}}(\theta);
            \end{equation}
            \begin{equation}
                |\rho_{2,5}(\theta)| < |\alpha|^{-1} \left \langle \frac{T_{\mathrm{PSF}}}{T_{\mathrm{gal}}} \right \rangle^{-1} \delta \xi_{+}^{\mathrm{max}}(\theta),
            \end{equation}
            where $\delta \xi_+^{\textrm{max}}$ is the sensitivity of $\xi_+$ with respect to cosmology. The PSF leakage $\alpha$ can be introduced as an additive bias in Eq.~\eqref{eq:def_e_obs}, leading to:
            \begin{equation}
            \label{eq:psf_leak}
                \left\langle e^{\mathrm{obs}}_{i} \right\rangle = (1+m)\; \left\langle \gamma_{i} \right\rangle + c_{i} + \alpha\; \left\langle e^{\mathrm{PSF}}_{i} \right\rangle.
            \end{equation}
            Following \cite{2017arXiv170801533Z}, we consider only one cosmological parameter here, namely $\sigma_8$, and get
            \begin{equation}
            \label{eq:rho_req}
                \delta \xi_{+}^{\mathrm{max}}(\theta) = \frac{\partial \xi_{+}(\theta)}{\partial \sigma_{8}} \delta \sigma_{8}.
            \end{equation}
            As shown in Eq.~(\eqref{eq:rho_req}), the requirements are defined with respect to $\sigma_{8}$. Since this paper does not make any claims on cosmology, this analysis is qualitative and we have set a soft requirement of 3\% error on $\sigma_{8}$, which means ${\delta\sigma_{8}}/{\sigma_{8}} < 0.03$ (based on \cite{2017arXiv170801533Z}). To estimate the theoretical shear-shear correlation we make use of the \textsc{CCL} software package\footnote{\url{https://github.com/LSSTDESC/CCL}} \citep{Chisari_2019} with the estimated $N(z)$ described in Sect.~\ref{sec:redshift_distribution}. The results are presented in Fig.~\ref{fig:rho_stat}. The PSF residuals are sub-dominant at scales smaller than around $100$\arcmin. We can see that the model performs worse on the largest scales. This might be related to the issues raised regarding the PSF residuals (Fig.~\ref{fig:psf_res}). The inconsistency of the model for large spatial variations is also reflected in these statistics.
        
    \subsection{Shear validation tests}
    
        This section presents the tests we perform on our shape catalogue to estimate the systematics on the data. First, we compute the additive shear bias. Next, we consider the residual correlation of galaxy shapes with the PSF, which is a main concern for shear estimation. To quantify this effect we carry out three tests, presented below. To perform our tests we use the definition for the PSF leakage, $\alpha$, presented in Eq.~\eqref{eq:psf_leak}.

        \subsubsection{Additive shear bias}
        
        To estimate the additive bias, $c_i$ defined in Eq.~\eqref{eq:def_e_obs}, we compute the weighted average of both components of the galaxy ellipticity using a jackknife estimator. Since our data is observed over a very large area, we can safely assume that the average shear and intrinsic ellipticity are very close to zero. With this assumption, we measure $c_1 = (-4.95 \pm 0.58) \times 10^{-4}$, and $c_2 = (4.66 \pm 0.59) \times 10^{-4}$ for the two ellipticity components. These numbers are of the same order of magnitude compared to previous measurements for this shape measurement method \citep{2017ApJ...841...24S}. It is important to note that a small additive bias does not imply zero PSF leakage, as we have observed.

        \subsubsection{Global PSF leakage}
        \label{sec:global_leakage}
        
\begin{figure*}
            \centering
                \resizebox{\hsize}{!}
                {
                    \begin{minipage}{.5\textwidth}
                      \centering
                      \includegraphics[width=1\linewidth]{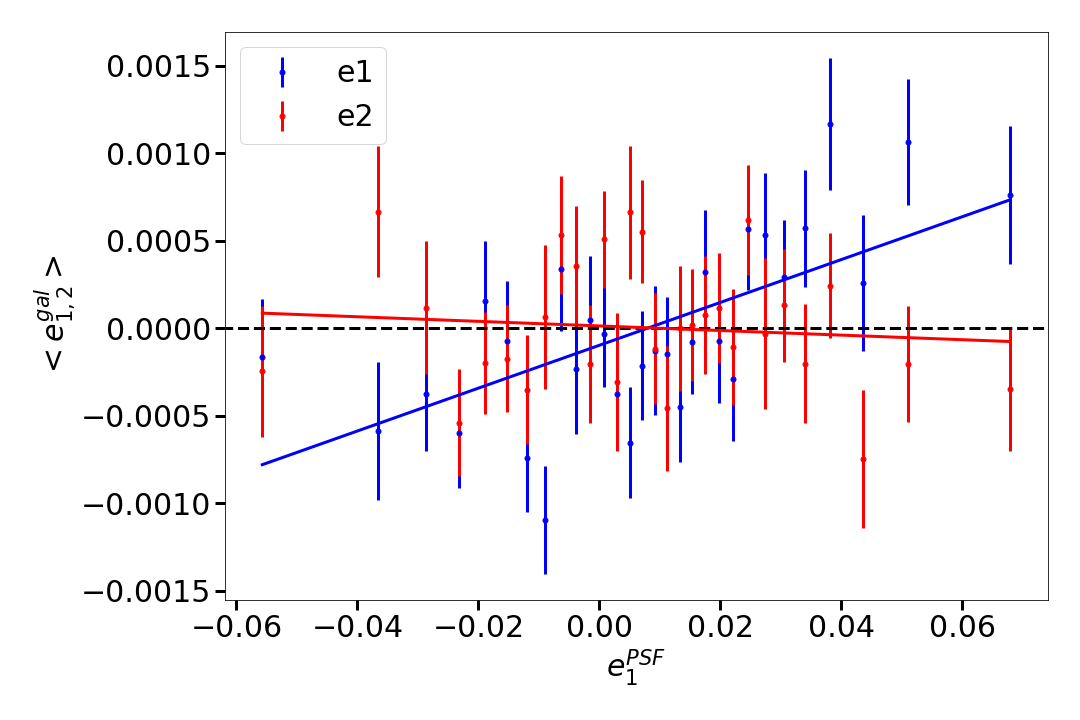}
                    \end{minipage}
                    \begin{minipage}{.5\textwidth}
                      \centering
                      \includegraphics[width=1\linewidth]{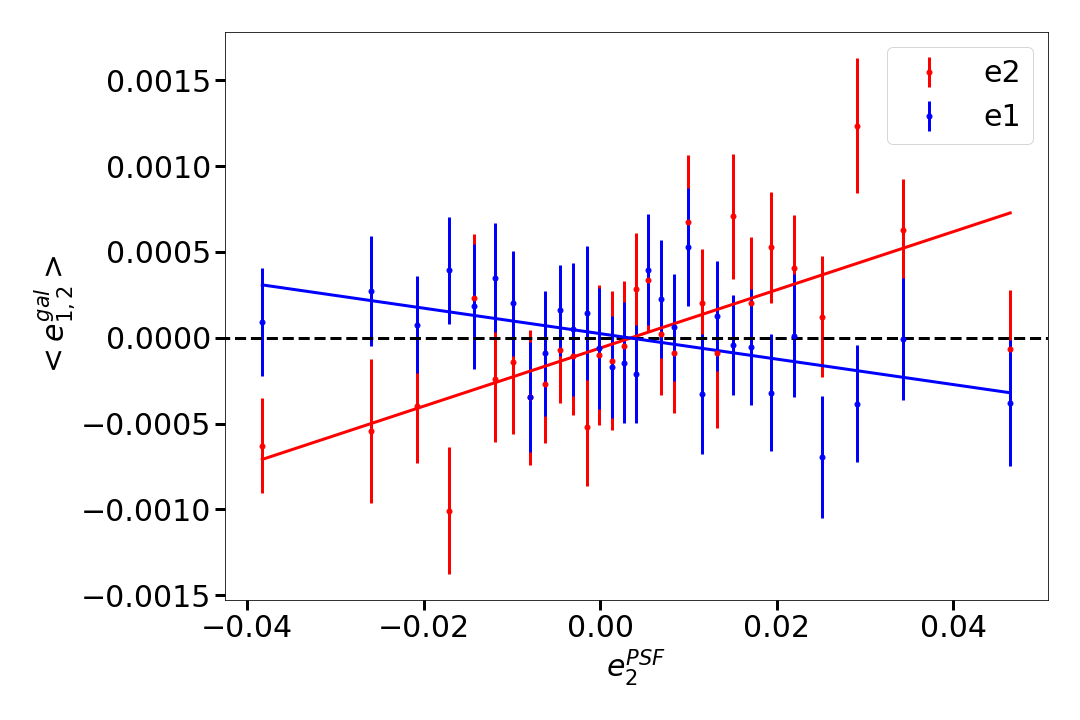}
                    \end{minipage}
                }
                \caption{ PSF leakage using the averaged galaxy shape in bins of PSF ellipticity component 1 (\textit{left panel}) and component 2 (\textit{right panel}). For the figure on the \textit{left}, we find $\langle e_{1}^{\mathrm{gal}} \rangle = (0.012 \pm 0.002)~<e_{1}^{\mathrm{PSF}}>$ and $\langle e_{2}^{\mathrm{gal}} \rangle = (-0.001 \pm 0.002)~<e_{1}^{\mathrm{PSF}}>$. For the figure on the \textit{right} we find $\langle e_{2}^{\mathrm{gal}} \rangle = (0.017 \pm 0.003)~<e_{2}^{\mathrm{PSF}}>$ and $\langle e_{1}^{\mathrm{gal}} \rangle = (-0.007 \pm 0.003)~<e_{2}^{\mathrm{PSF}}>$.}
                \label{fig:PSF_leakage_ell}
                \end{figure*}

\begin{figure}[h!]
                \centering
                	\includegraphics[width=1.\linewidth]{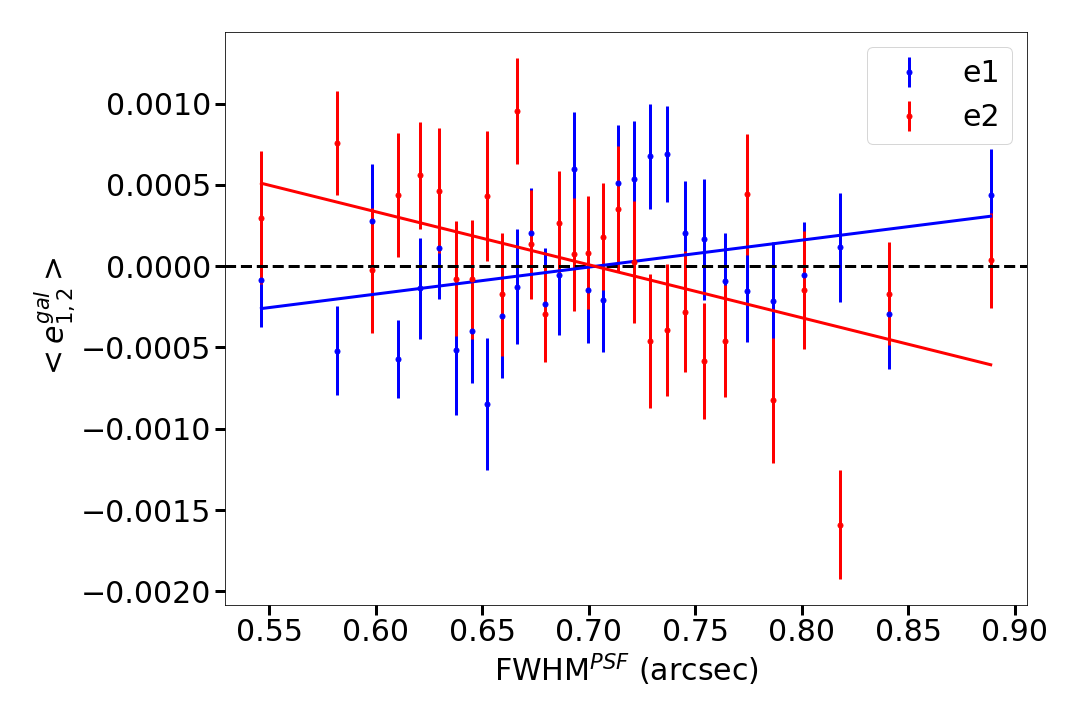}
                	\caption{PSF leakage using the averaged galaxy shape in bins of PSF size. The slopes are $\langle e_{1}^{\mathrm{gal}} \rangle = (0.002 \pm 0.0007)~<\mathrm{FWHM}^{\mathrm{PSF}}>$ and $\langle e_{2}^{\mathrm{gal}} \rangle = (-0.003 \pm 0.0007)~<\mathrm{FWHM}^{\mathrm{PSF}}>$.}
                	\label{fig:PSF_leakage_size}
                \end{figure}

            In the first test, for which the results are shown in Fig.~\ref{fig:PSF_leakage_ell}, we estimate the global PSF leakage by measuring the galaxy ellipticies in bins of PSF ellipticity. For this test we use the  PSF ellipticity measured at the position of galaxies averaged over the contributing single exposures. For each of the equi-populated bins we estimate the weighted average via a jackknife, where the error bars represent the standard deviation. The weights are defined in Eq.~(\eqref{eq:ell_weight}). We find correlations between $e^{\mathrm{gal}}_i$ and $e^{\mathrm{PSF}}_i$ of less than $2\%$ for both components $i=1, 2$. The cross-correlation between different components is close to $0$. No correlations are observed with the PSF size as presented in Fig.~\ref{fig:PSF_leakage_size}. To compute the correlations we fit a linear model on the unbinned data.
        
        \subsubsection{Scale-dependent PSF leakage}
        \label{sec:PSF_leakage}
        
\begin{figure}[h!]
                \centering
                	\includegraphics[width=1.\linewidth]{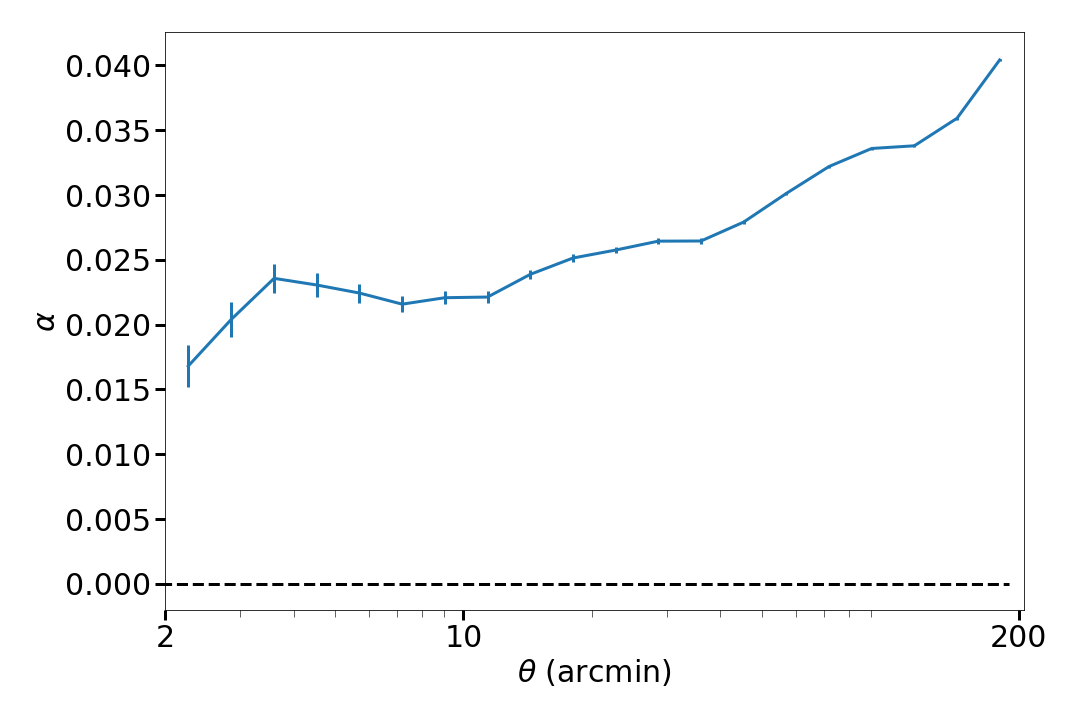}
                	\caption{PSF leakage $\alpha$ (Eq.~\eqref{eq:alpha_scale}) shown as function of scale $\theta$.}
                	\label{fig:alpha_leakage}
            \end{figure}

            Another test we perform to estimate the leakage was presented in \cite{2016MNRAS.460.2245J}. The leakage $\alpha$ can be written as the ratio between the star-galaxy cross-correlation, $\xi^{\rm gp}_+$, and the star-star auto-correlation, $\xi^{\rm pp}_+$. For this test, we use the ellipticity of the PSF model at the position of the stars (test sample only).
            It is defined as follows:
            \begin{equation}
                \alpha(\theta) = \frac{\xi_{+}^{\mathrm{gp}}(\theta) - \langle e_{\mathrm{gal}} \rangle^{*} \langle e_{\mathrm{PSF}} \rangle}{\xi_{+}^{\mathrm{pp}}(\theta) - |\langle e_{\mathrm{PSF}} \rangle|^{2}}.
                \label{eq:alpha_scale}
            \end{equation}
            The variations of $\alpha$ in the range $[2, 200]$ arcmin is presented in Fig.~\ref{fig:alpha_leakage}. The all-scale average gives $\alpha = 0.033$ for the model-fitting method. $\alpha$ is in agreement with the test presented above when $\theta \rightarrow 0$.
            
\begin{figure}[h!]
                \centering
                	\includegraphics[width=1.\linewidth]{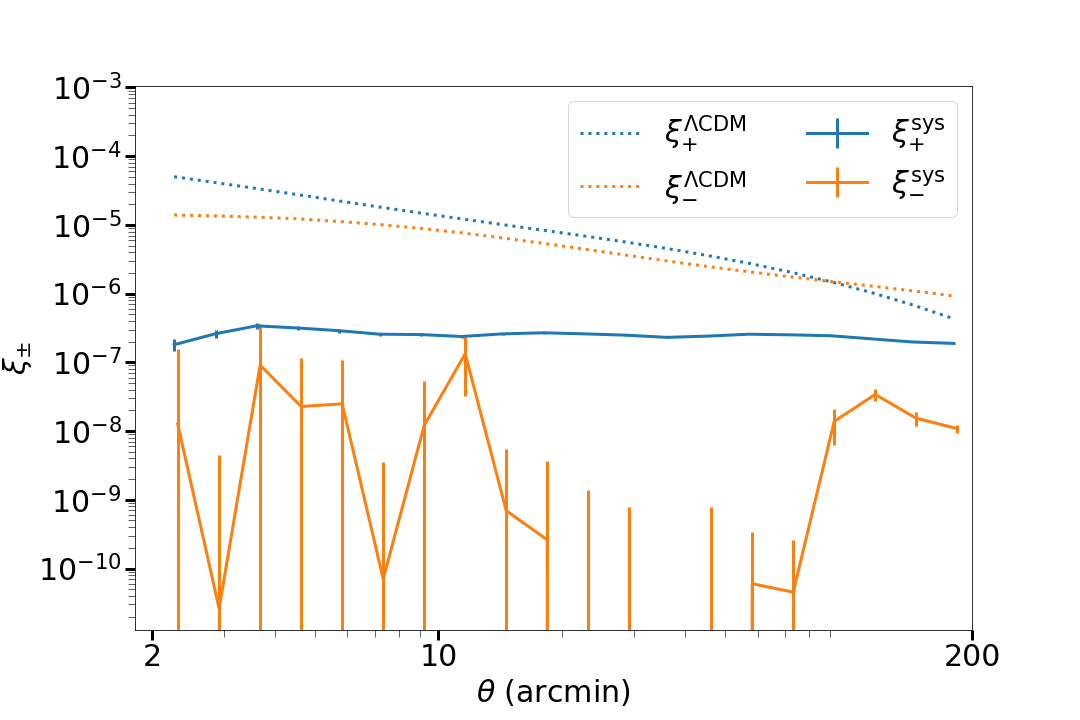}
                	\caption{Systematic signal $\xi^{\mathrm{sys}}_{\pm}$ (Eq.~\eqref{eq:xi_sys}) compared to the theoretical shear-shear correlations.}
                	\label{fig:xi_sys}
            \end{figure}

            Finally, we look at  $\xi^{\rm sys}$, which was introduced in \cite{2003MNRAS.344..673B}, and has been used in \cite{CFHTLenS-sys} for the CFHTLenS survey. This consists in comparing the signal of the shear-PSF correlation to the shear-shear correlation. If the PSF is well corrected for, the former (systematic) has an amplitude that is much lower than the latter (signal), such that it does not contaminate the cosmological interpretation of the latter. The systematic correlation function can be written as:
            \begin{equation}
            \label{eq:xi_sys}
                \xi^{\mathrm{sys}}_{\pm}(\theta) =
                \frac{\left(\xi_{\pm}^{\mathrm{gp}}
                    \right)^2(\theta)}
                    {\xi_{\pm}^{\mathrm{pp}}(\theta)} .
            \end{equation}
            Results are shown in Fig.~\ref{fig:xi_sys}. To see the impact of our systematics we used a theoretical prediction for the shear-shear correlations, $\xi_{+}^{\mathrm{\Lambda CDM}}$, using the \textsc{CCL} library and cosmological parameters from \textit{Planck} \citep{2018arXiv180706209P}. We see that on small scales we have a signal \editt{\sout{virtually}} free from systematics, which are two orders of magnitudes smaller. However, on large scales $\xi^{\mathrm{sys}}$ reaches $10$--$20$\% of the shear-shear correlation, and contributions from residual PSF correlations to the cosmological signal might not be negligible.
            
            Overall, the different tests performed on our shape catalogue show very small correlations with the PSF of the order of a few percent. The large scales are still impacted by our PSF modelling. We believe that this problem can be solved by having a more accurate PSF model at all scales. For the purposes of the work presented here we have neglected the impact of blends, which will be the main concern for the future development of the weak-lensing analysis of CFIS data.


\section{Image simulations}
\label{sec:simulation}

    In this section we present the set of image simulations we use to validate the shape measurements. The simulations have been created to test the implementation of the model-fitting method. We implement all effects that need to be accounted for during the shape measurement, that is: transformations between a planar and spherical world coordinate system (WCS), miss-centering, spatial noise and PSF variation. We have used the simulations to quantify the PSF leakage, and the additive and multiplicative biases. Only the multiplicative bias will be detailed here since it is not available from the data only. The other tests show results consistent with those obtained from the data.
    
    \subsection{Simulated PSF}
    
        The PSF can be separated in two parts: the optical distortions and the atmospheric turbulence. To simulate the optical variations of the telescope we use a Moffat profile for which we fix the atmospheric scattering coefficient $\beta=4.765$ \citep{beta_moffat_stars}, and the ellipticity is drawn from the real optical variations of CFHT (Canada-France-Hawaii Telescope) derived from the CFIS data shown in Fig.~\ref{fig:psf_res}. To model the atmospheric turbulence we use a Kolmogorov profile \citep{kolmogorov_model} with random ellipticity drawn from a Gaussian distribution with mean $\mu = 0$ and standard deviation $\sigma=0.01$. Both models are convolved to create the final PSF. The average size of the simulated PSFs is set to 0.65~arcsec, which corresponds to the mean seeing of CFIS (see Sect.~\ref{sec:star_selection}). This model is a simple but accurate enough description of the real PSF obtained from the data. Since we are working on small postage stamps, the spatial variations across the postage stamp have been neglected. This process is repeated for each simulated observation of each object. This allows us to quantify the PSF leakage in the simulations. We also applied WCS transformations to the PSF used for the shape measurement, which were randomly selected among the real images.
    
    \subsection{Real galaxies}
    
        Our set of simulations is based on the real galaxy images from the COSMOS catalogue. Here, the deconvolved images are used\footnote{This step creates artefacts and correlated noise residuals. However, after reconvolution with a larger PSF (described above), and with the addition of noise, it can be considered that these spurious correlations have been removed from the final image.}. The flux is rescaled in order to reproduce a 300s exposure at the 3.6m CFHT telescope. The image is resampled at the pixel scale of the CFIS survey, 0.187 arcsec. Finally, Gaussian noise with $\sigma = 14.5$ is added on top of the image to replicate the SNR on the CFIS data. We also include WCS transformations on the images by random draws from the real data. We create 3 epochs of observations for each object. Between epochs we vary the centering (intra-pixel shifts), the noise realisation, and the PSF.
        
        The galaxies are created in batches of $10{,}000$ postage stamps with the same constant shear applied to them. One half of the galaxies are copies of the other half but rotated by 90\degree\, to cancel out shape noise. The rotation is applied before the shear is added. We simulate $200$ batches  with $200$ different shear values.

    \subsection{Shear bias estimation}
    
\begin{figure}[h!]
        	\centering
        	\includegraphics[width=1.\linewidth]{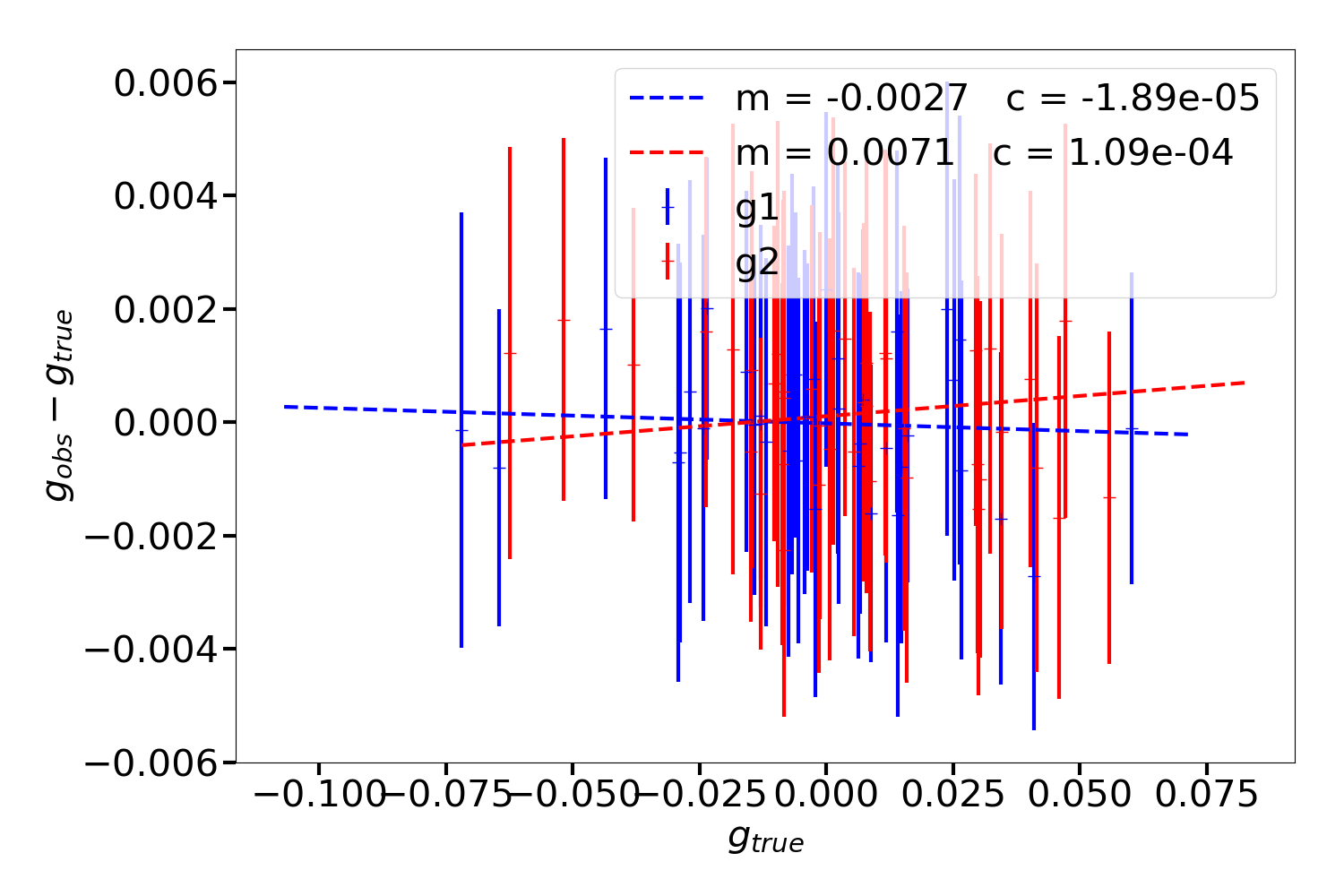}
        	\caption{Residual multiplicative bias after calibration and weighting. \textit{Dashed} lines represent the best linear fit ($m~g_{\mathrm{true}} + c$) to the points.}
        	\label{fig:simu_mul_bias}
        \end{figure}

        The main purpose of our image simulations is to quantify the residual shear bias after calibration. For this reason, only the shape measurement is performed. We neglect all biases coming from detection or the pre-selection cuts. The results are presented in Fig.~\ref{fig:simu_mul_bias}, showing that, after calibration, we are left with a residual multiplicative bias of the order $10^{-3}$, and an additive bias of at most $10^{-4}$.
        
        As mentioned previously, we can use the image simulations to perform an independent consistency check of the PSF leakage created by our analysis. Here we measure the global PSF leakage presented in Sect.~\ref{sec:global_leakage}. We find a leakage compatible with zero for both components of the ellipticity, as shown in Figs.~\ref{fig:PSF_leakage_simu_ell} and \ref{fig:PSF_leakage_simu_size}. Since the PSF in the image simulations is perfectly known, this is a strong indication that the leakage stems from an imperfect modelling of the PSF variation across the focal plane. This has been discussed in \cite{2017arXiv170801533Z}.

\section{Science results}
\label{sec:science}

    In this section we present the first science results obtained with our shear catalogue. This section serves as a proof of concept of the results achievable in the future with the full CFIS data and processing with our pipeline, \textsc{ShapePipe}. We focus here on convergence maps, cluster lensing, and second-order cosmic shear statistics.
    
    \subsection{Shear catalogue}
    
\begin{figure}[h!]
        	\centering
        	\includegraphics[width=1.\linewidth]{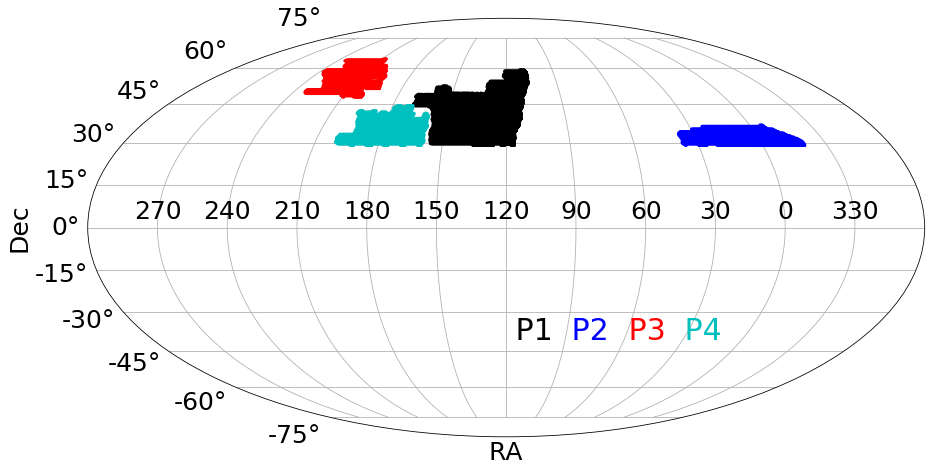}
        	\caption{The four patches of the CFIS dataset processed in this work. From left to right: P3, P4, P1 and P2.}
        	\label{fig:patch_footprint}
        \end{figure}

\begin{figure}[h!]
        \centering
        	\includegraphics[width=1.\linewidth]{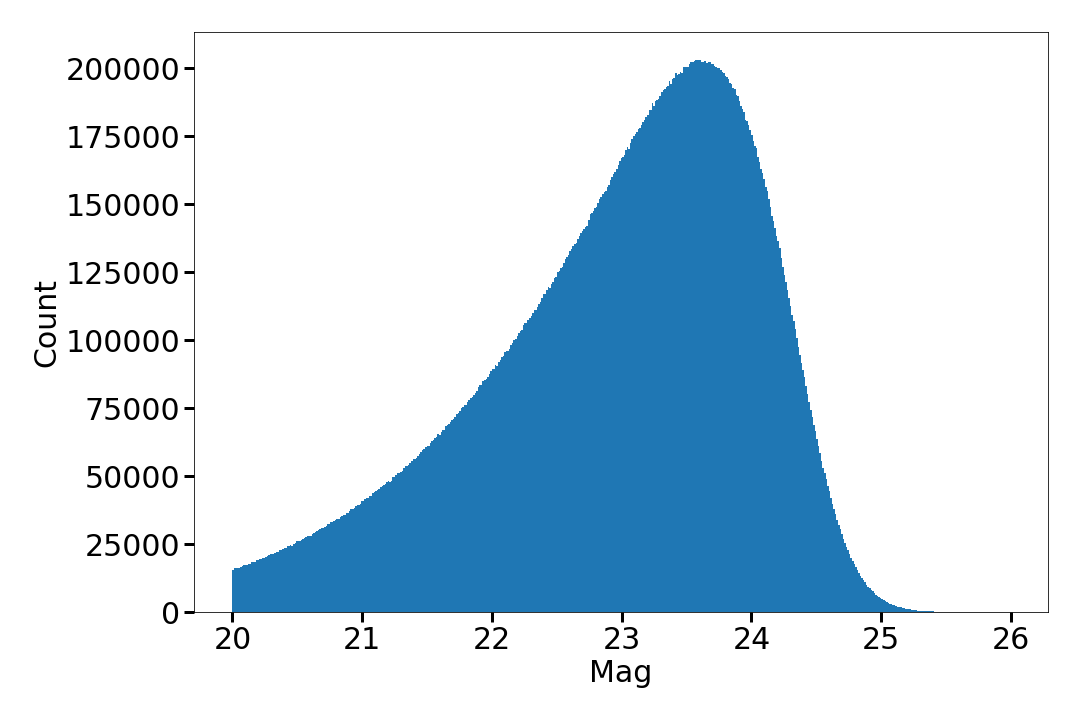}
        	\caption{Magnitude distribution on the $r$-band for weak-lensing selected source galaxies.}
        	\label{fig:mag_distribution}
        \end{figure}

         The catalogue used here is constructed from an effective (after masking) area of $A = 1\,565$ square degrees. We measure the shape of $N_{\mathrm{gal}} = 40\,151\,119$ galaxies. The data are divided in 4 fields, named P1 to P4, which are presented in Fig.~\ref{fig:patch_footprint}. The effective density is derived from the formula proposed in \cite{CFHTLenS-sys}:
         \begin{equation}
             n_{\mathrm{eff}} = \frac{1}{A} \frac{\left( \sum w_{i} \right)^{2}}{\sum w^{2}_{i}},
         \end{equation}
         and the corresponding shape noise is given by:
         \begin{equation}
         \label{eq:shape_noise}
             \sigma^{2}_{\mathrm{SN}} = \frac{1}{2} \frac{\sum w^{2}_{i} (e^{2}_{1,i} + e^{2}_{2,i})}{\sum w^{2}_{i}},
         \end{equation}
         where the sum is carried out  over all objects in the catalogue. We find $n_{\mathrm{eff}} = 6.76~\mathrm{gal} \, \mathrm{arcmin}^{-2}$, compared to a raw, unweighted density of $n_{\mathrm{raw}} =  N_{\mathrm{gal}} / A =  7.13~\mathrm{gal}\, \mathrm{arcmin}^{-2}$, and $\sigma_{\mathrm{SN}} = 0.35$. The magnitude distribution is presented in Fig.~\ref{fig:mag_distribution}. Due to our conservative detection criteria, this distribution does not reflect the true depth of the survey. With a selection aimed at completeness, we should be able to obtain a complete sample at magnitude 23.5 in the $r$-band.
         
    \subsection{Convergence mass maps}
    \label{sec:mass_map}
    
        Convergence maps can be interpreted as 2D projected distributions of matter fluctuations. To compute the convergence from the shear catalogue we perform a Kaiser-Squires inversion \citep{1993ApJ...404..441K} using the \textsc{LensPack} software package\footnote{\url{https://github.com/CosmoStat/lenspack}}. The ellipticities are binned in $40\times40~\mathrm{arcsec}^{2}$ pixels. The masked regions have an $\epsilon$ set to 0. The convergence map is then smoothed by applying a Gaussian kernel with a standard deviation of 16 pixels.
        
        In Fig.~\ref{fig:mass_map_P1} we show patch P1, the other three patches are presented in Appendix~\ref{sec:annexe_mass_map}. Overplotted are the positions of known clusters detected via the Sunyaev-Zel'dovich (SZ) effect \citep{SZ_effect1,SZ_effect2}, presented in \cite{planck_clusters}.
        
\begin{figure*}
            \centering
            \hspace{-18.0mm}
            \includegraphics[width=1\textwidth]{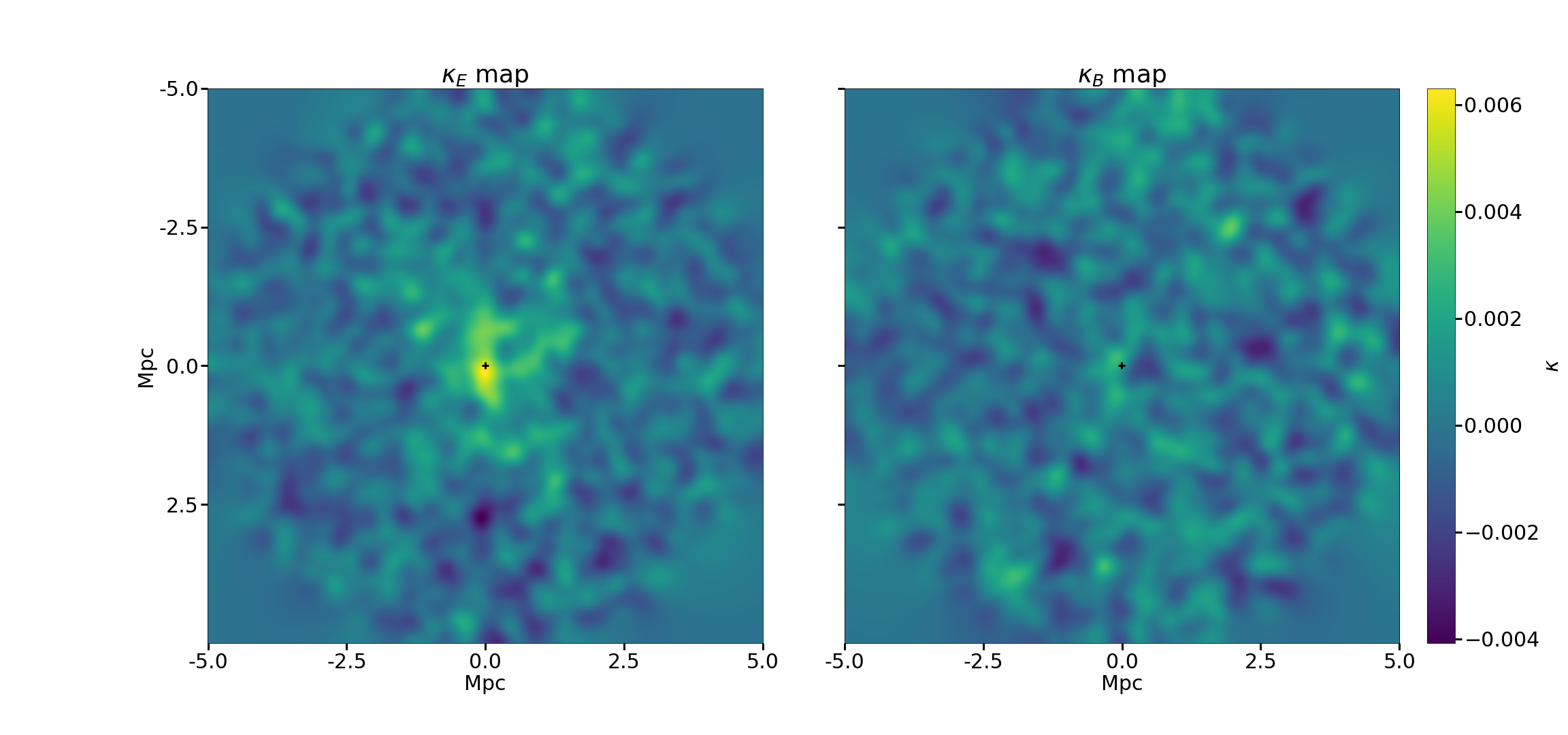}
            \caption{Mass maps stacked on the Planck cluster positions for P1. The galaxies for the tangential shear stacks are selected in a radius of $5~\mathrm{Mpc}$ around each cluster, where this distance is computed at the cluster redshift.}
            \label{fig:stacked_mass_map}
        \end{figure*}

        In the E-mode map, we see that most of the clusters match with an overdense region. On the other hand, in the B-mode, we do not see a correlation between the peaks and the cluster positions, as expected. To confirm this result, we show the convergence E- and B-mode maps stacked around the cluster positions. For this stack, we select source galaxies in a $5$ Mpc projected radius around each cluster. We then shift all galaxies to have the centre of every cluster at the same effective position, and we compute the Kaiser-Squires inversion from the joint galaxy ellipticities. The resulting maps are shown in Fig.~\ref{fig:stacked_mass_map}. In this plot we can clearly see that we obtain a strong signal in the E-mode, and only noise for the B-mode.
        
        \begin{figure*}
    		\centering
    		\vspace{-15.0mm}
    		\begin{minipage}{\textwidth}
    			\centering
    			\includegraphics[width=0.8\textwidth]{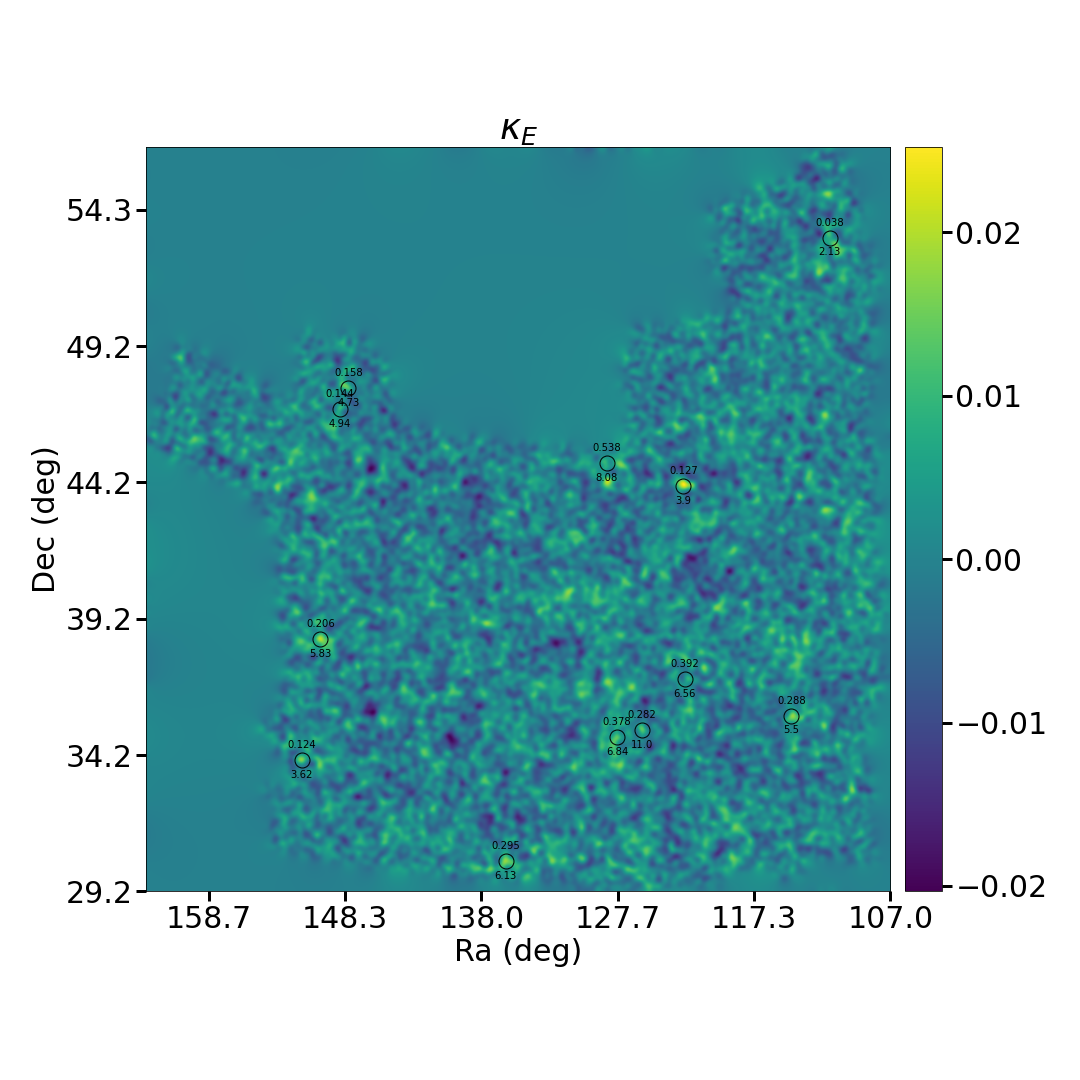}
    		\end{minipage}
    		\vspace{-30.0mm}
    		
    		\begin{minipage}{\textwidth}
    			\centering 
    			\includegraphics[width=0.8\textwidth]{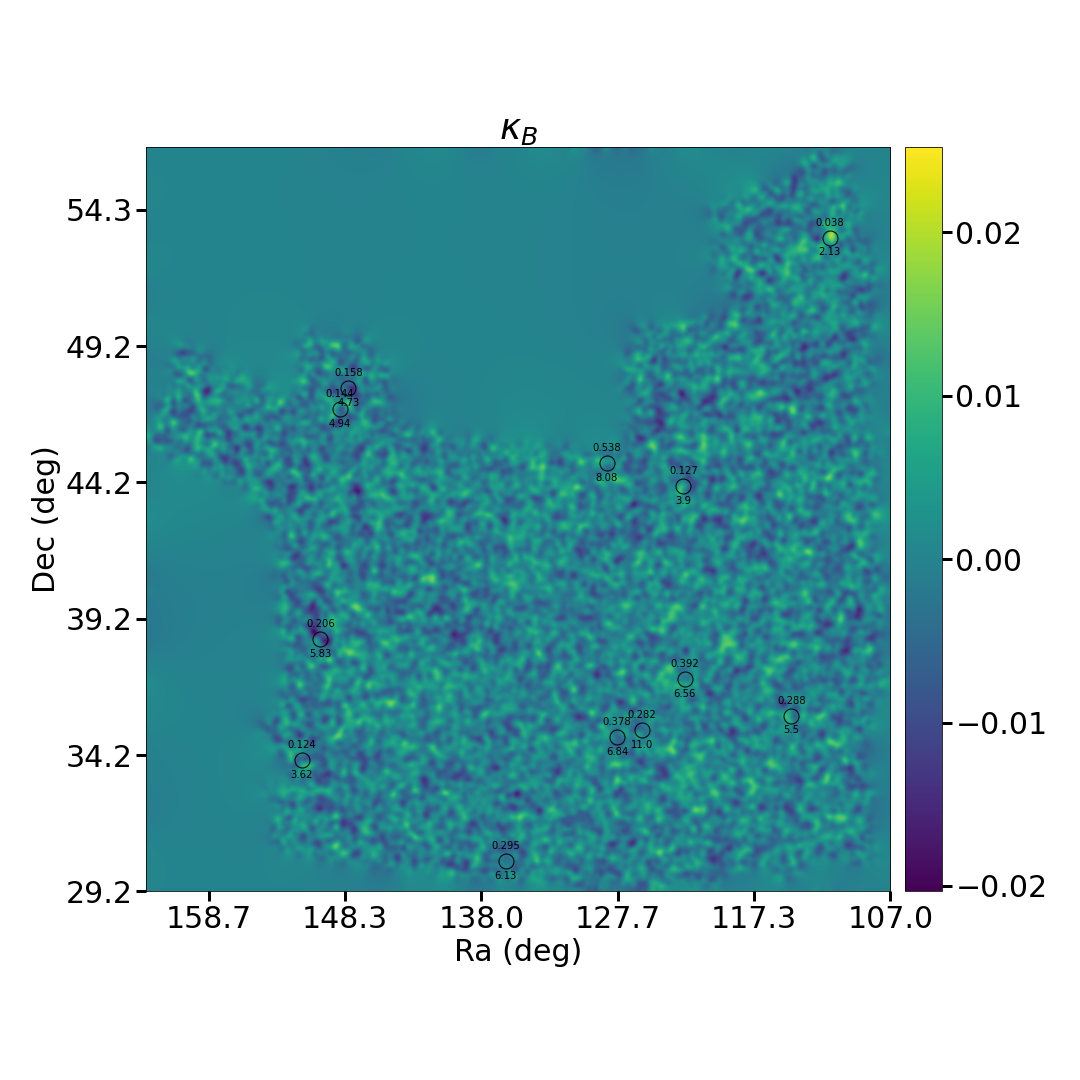}
    		\end{minipage}
    		\vspace{-15.0mm}
    
    		\caption{Mass map for the patch P1. The \textit{black circles} represent the positions of Planck clusters. The value on \textit{top} of each cross is the cluster redshift, and the \textit{bottom} value indicates the SZ cluster mass ($10^{14}\mathrm{M}_{\odot}$). The \textit{top} (\textit{bottom}) panel shows the E-mode (B-mode).}
    		\label{fig:mass_map_P1}
    	\end{figure*}
    	
    \subsection{Galaxy cluster stacked profiles}
    
\begin{figure}
            \centering
            \includegraphics[width=0.5\textwidth]{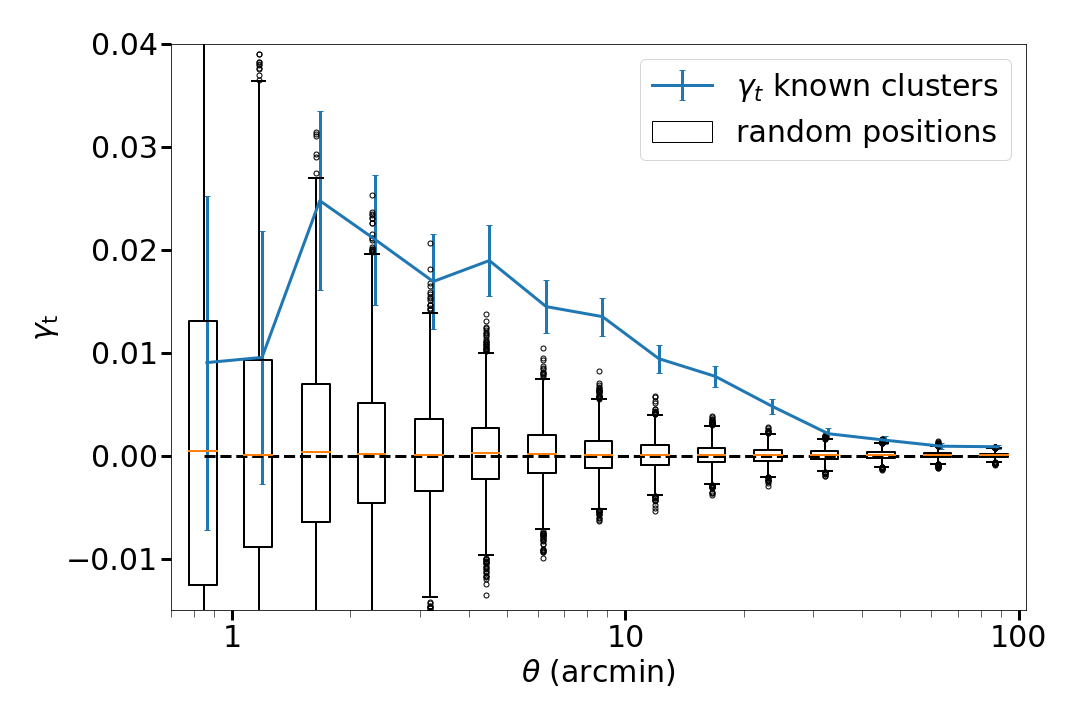}
            \caption{Stacked tangential shear around 36 clusters in the CFIS footprint. The box plot represents the result around $5000\times36$ random points.}
            \label{fig:cluster_bp}
        \end{figure}

        To evaluate the performance of our pipeline for future cluster science, we have measured the tangential shear profile around the clusters used in the previous section (36 clusters fall within the processed area after masking). To assess the significance of the signal we compare it to what we measure at random positions. The box plot presented in Fig.~\ref{fig:cluster_bp} is constructed from random catalogues. We create catalogues that have the same size as the real one (36 here) with random positions. We  repeat this process $5\,000$ times. We can then derive the variance and the median of the tangential shear around these random positions. In this figure we see that the random positions produce a signal compatible with 0. In comparison, the tangential shear signal around the true cluster positions is detected at a high significance. The tangential shear at a given angular scale exceeds the random signal on average by $4~\sigma$, which can be seen as a conservative, lower bound of the overall detection significance.
        
        In a separate analysis, we (Spitzer et al., in prep.) have analyzed the weak-lensing masses of redMapper clusters \citep{2014ApJ...785..104R} using this ShapePipe catalogue and find good agreement with previous results from weak-lensing with SDSS \citep{2017MNRAS.466.3103S}.
        
    \subsection{Second-order shear statistics: COSEBI B-modes}
    \label{sec:cosebis}
    
        One of the main weak-lensing observables for cosmology are the second-order cosmic shear correlations. One such measure are the Complete Orthogonal Sets of E-/B-mode Integrals (COSEBIs), introduced in \cite{COSEBIs}. These modes are linear functions of the measured two-point correlation functions $\xi_+$ and $\xi_-$ (Sect.~\ref{sec:2pcf}).
        
        First, we compute the shear-shear correlation functions between $\theta_{\mathrm{min}}=2~\mathrm{arcmin}$ and $\theta_{\mathrm{max}}=200~\mathrm{arcmin}$ with $1000~\mathrm{bins}$ using the \textsc{treecorr}\footnote{\url{https://github.com/rmjarvis/TreeCorr}} software package \citep{treecorr_cite}. Next, we derive  the COSEBIs using \textsc{nicaea}\footnote{\url{https://github.com/CosmoStat/nicaea}} \citep{nicaea_cite}.
        This results in E- and B-modes, $E_m$ and $B_m$, respectively, where each of these modes is an integral over $\xi_+(\theta)$ and $\xi_-(\theta)$ filtered with a polynomial in $\log \theta$ of order $m+1$. 
        
\begin{figure}
            \centering
            \includegraphics[width=0.5\textwidth]{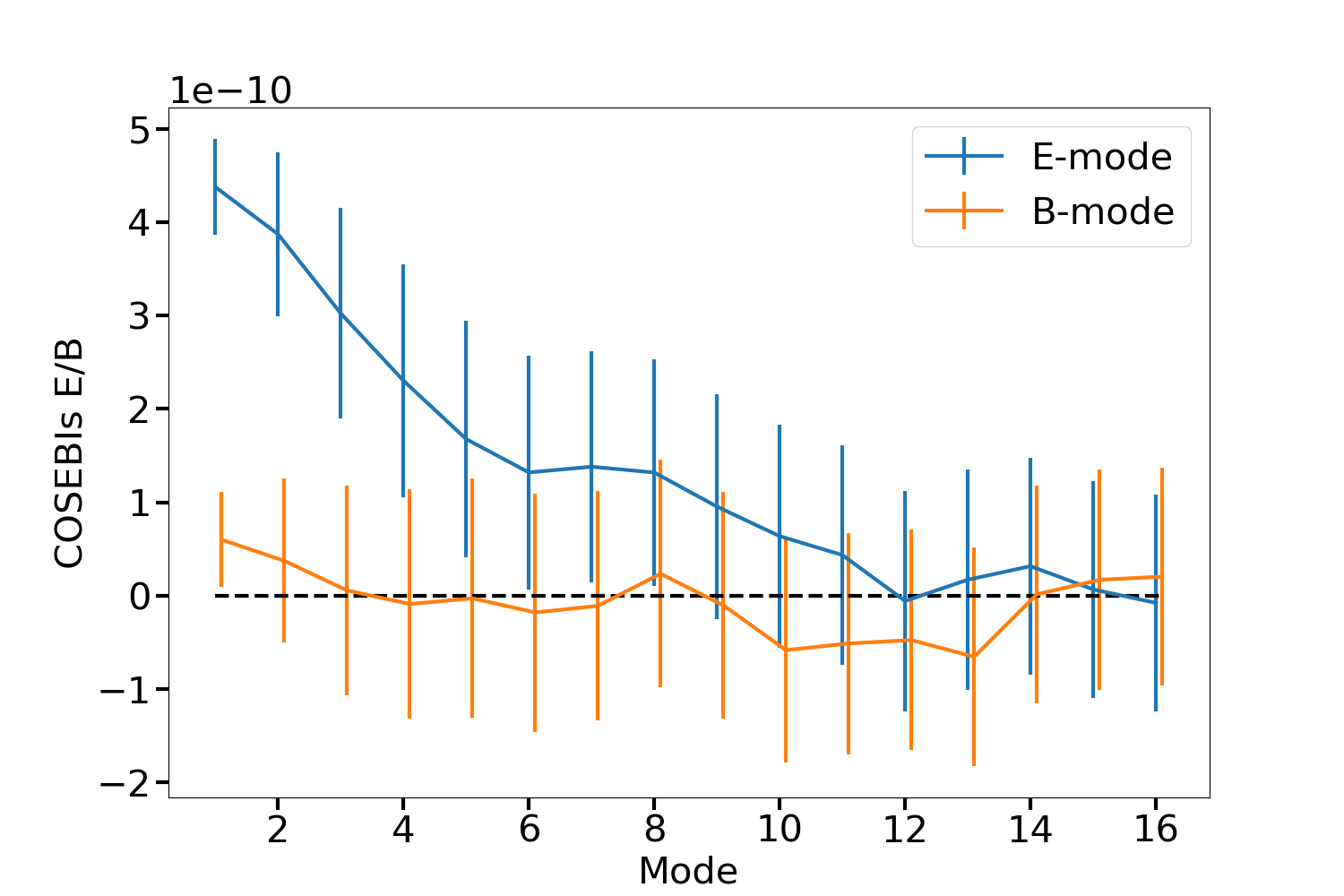}
            \caption{We show here the COSEBIs for modes $m$ between 1 and 16. This test was performed to quantify the impact of potential systematics on the B-mode, which are consistent with 0 for our study.}
            \label{fig:cosebis}
        \end{figure}

        To construct the error bars, we first assume a Gaussian covariance for $\xi_{\pm}$, derived from the \textsc{CosmoCov} software package\footnote{\url{https://github.com/CosmoLike/CosmoCov}} (\cite{cosmocov_cite1}, \citeauthor{cosmocov_cite2} (\citeyear{cosmocov_cite2}, \citeyear{cosmocov_cite3})). 
        As a fiducial model for the covariance we use a \textit{Planck} cosmology \citep{2018arXiv180706209P}, and ignore intrinsic galaxy alignment.
        Next, we re-sample the shear-shear correlations $1\,000$ times from this covariance, and re-compute the COSEBIs for each simulated $\xi_{\pm}$. The results are presented in Fig.~\ref{fig:cosebis}. We see that the B-modes are consistent with 0, which indicates a low level of systematics present at second order in our shear data.
        
\section{Conclusion}
\label{sec:conclusion}

    In this paper we have introduced a new weak-lensing pipeline, \ShapePipe. This pipeline relies on well-tested tools such as \textsc{SExtractor}, \textsc{PSFEx}, and \textsc{ngmix}. These tools have been adapted to suit the demanding constraints on memory and CPU time imposed by the large area of the CFIS survey. We have demonstrated that our pipeline is able to handle such large data sets. The modularity of the pipeline makes it easy to adapt it to new data sets, and to update it to include new  state-of-the-art shape measurement and analysis tools.
    
    This first analysis of $1\,700$ deg$^2$ of CFIS data presented in this work shows the high quality of the images as well as the quality of both pre-processing and processing. The most well-known systematics that create biases in the shear estimated from galaxy shape measurements have been addressed in this work. Our modeling of the PSF captures the measured PSF accurately, with typical residuals an order of magnitude smaller than the focal-plane PSF variations, and showing remaining coherent patterns at only a very small amplitude.
    The PSF residual angular correlations ($\rho$ statistics) are sub-dominant to the shear two-point correlation function on scales up to $100$\arcmin. This corresponds to a sub-dominant contribution to a potential measurement of $\sigma_8$ to $3\%$ accuracy.
    
    We also have quantified the contamination of galaxy ellipticities by the PSF shape, the so-called PSF leakage. The global leakage is below $2\%$, which we measure using two different methods. For larger angular distances the scale-dependent leakage increases to $4\%$. Through comparison with galaxy shapes measured on simulations, we have established this leakage is caused by the remaining PSF residuals.
    
    Our preliminary science analysis shows that the pipeline is well suited for cluster lensing or mass mapping science. The COSEBIs confirmed the low level of systematics for shear-shear correlations observed on our different validation tests.
    
    Our main remaining source of systematic seems to come from the PSF modeling. In a future analysis we will employ a new approach based on the work from \cite{Liaudat_2021}, with the objective to reduce the residuals at large scales. Finally, we have ignored the effect of blending for this first analysis. This will be addressed in more detail in future works.


\begin{acknowledgements}

We would like to thank Catherine Heymans, Mike Jarvis, François Lanusse, Erin Sheldon, and Florent Sureau for helpful discussions.

This work is based on data obtained as part of the Canada- France Imaging Survey, a CFHT large program of the National Research Council of Canada and the French Centre National de la Recherche Scientifique. Based on observations obtained with MegaPrime/MegaCam, a joint project of CFHT and CEA Saclay, at the Canada-France-Hawaii Telescope (CFHT) which is operated by the National Research Council (NRC) of Canada, the Institut National des Science de l’Univers (INSU) of the Centre National de la Recherche Scientifique (CNRS) of France, and the University of Hawaii.
Pan-STARRS is a project of the Institute for Astronomy of the University of Hawaii, and is supported by the NASA SSO Near Earth Observation Program under grants 80NSSC18K0971, NNX14AM74G, NNX12AR65G, NNX13AQ47G, NNX08AR22G, and by the State of Hawaii.

We also acknowledge the support
from the French national program for cosmology and galaxies (PNCG).
%
This work is supported by R\'egion d’\^Ile-de-France in the
framework of DIM-ACAV thesis fellowship.
This work was supported in part by the Canadian Advanced Network for Astronomical Research (CANFAR) and Compute Canada facilities.
This work has made use of the CANDIDE Cluster at the Institut d'Astrophysique de Paris and made possible by grants from the PNCG and the DIM-ACAV.
We gratefully acknowledge support from the CNRS/IN2P3 Computing Center (Lyon -
France) for providing computing and data-processing resources.
This research made use of
Astropy\footnote{\url{http://www.astropy.org}}, a community-developed core Python
package for Astronomy \citep{astropy:2013, astropy:2018}.
Arnau Pujol acknowledges support from a European Research Council Starting Grant (LENA-678282) and Juan de la Cierva fellowship.
Hendrik Hildebrandt is supported by a Heisenberg grant of the Deutsche Forschungsgemeinschaft (Hi 1495/5-1) as well as an ERC Consolidator Grant (No. 770935).
Raphael Gavazzi thanks IoA and the Churchill College in Cambridge for their 
hospitality and acknowledges local support from the French government.
Mike Hudson acknowledges support from an NSERC Discovery Grant.
\end{acknowledgements}

\bibliographystyle{aa}
\bibliography{astro}


\begin{appendix}

\section{Leakage PSF in simulation}
\label{sec:annexe_psf_leakage}

\begin{figure*}
    \centering
        \resizebox{\hsize}{!}
        {
            \begin{minipage}{.5\textwidth}
              \centering
              \includegraphics[width=1\linewidth]{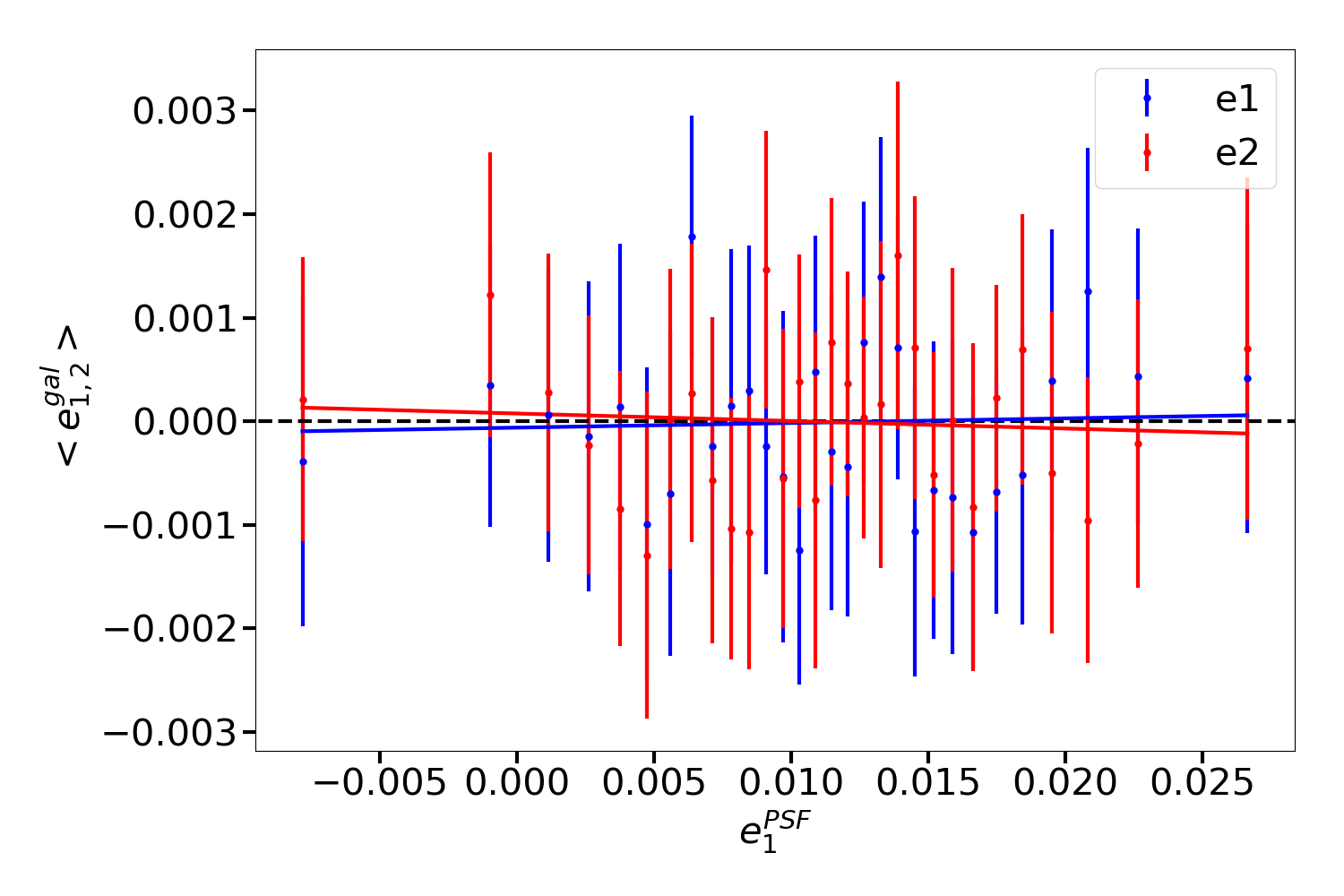}
            \end{minipage}
            \begin{minipage}{.5\textwidth}
              \centering
              \includegraphics[width=1\linewidth]{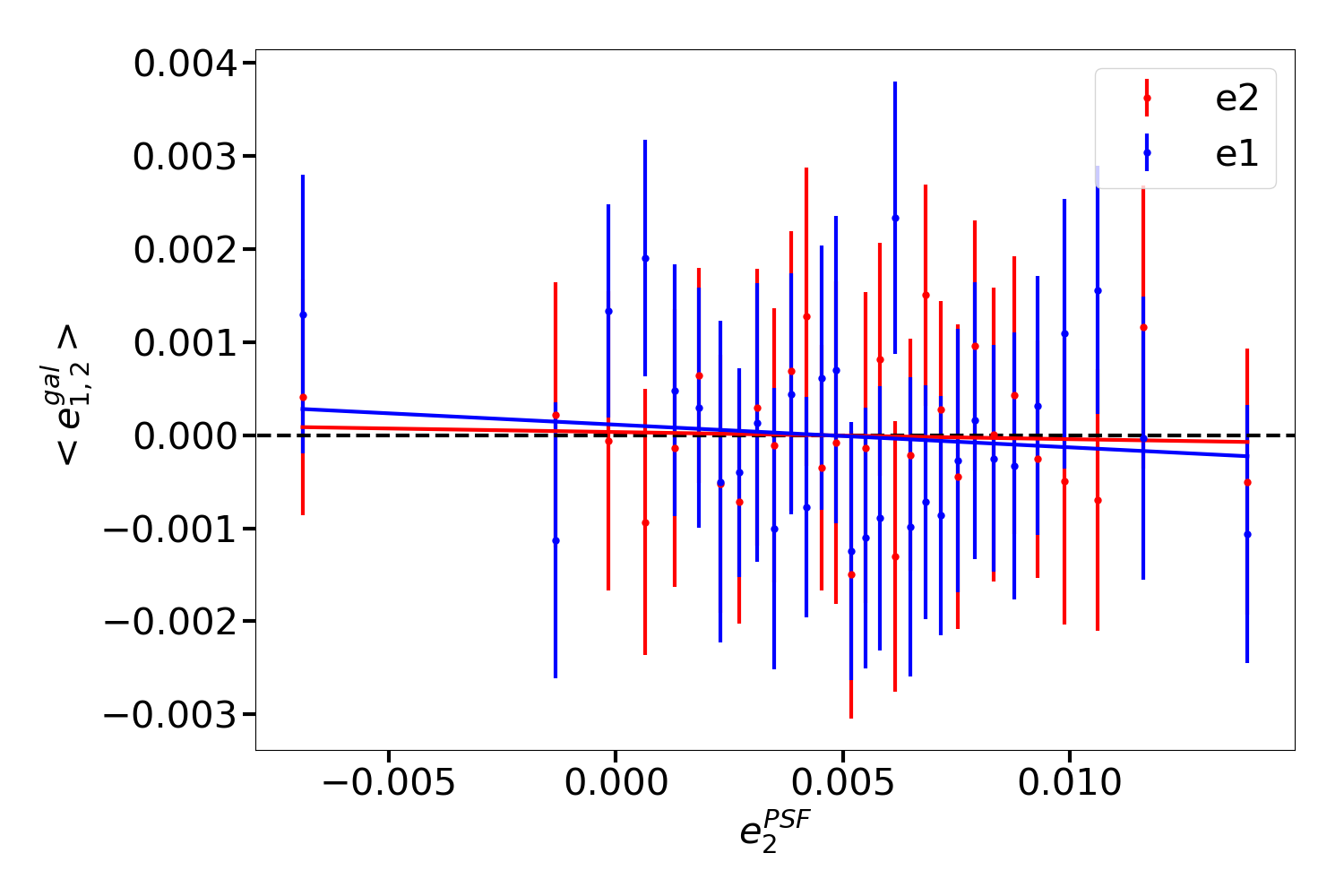}
            \end{minipage}
        }
        \caption{PSF leakage using the averaged galaxy shape in bins of PSF ellipticity component 1 (\textit{left panel}) and component 2 (\textit{right panel}). For the figure on the \textit{left}, we find $\langle e_{1}^{\mathrm{gal}} \rangle = (0.004 \pm 0.03)~<e_{1}^{\mathrm{PSF}}>$ and $\langle e_{2}^{\mathrm{gal}} \rangle = (-0.007 \pm 0.03)~<e_{1}^{\mathrm{PSF}}>$. For the figure on the \textit{right} we find $\langle e_{2}^{\mathrm{gal}} \rangle = (-0.008 \pm 0.05)~<e_{2}^{\mathrm{PSF}}>$ and $\langle e_{1}^{\mathrm{gal}} \rangle = (-0.024 \pm 0.05)~<e_{2}^{\mathrm{PSF}}>$.}
        \label{fig:PSF_leakage_simu_ell}
        \end{figure*}

\begin{figure}[h!]
        \centering
        	\includegraphics[width=1.\linewidth]{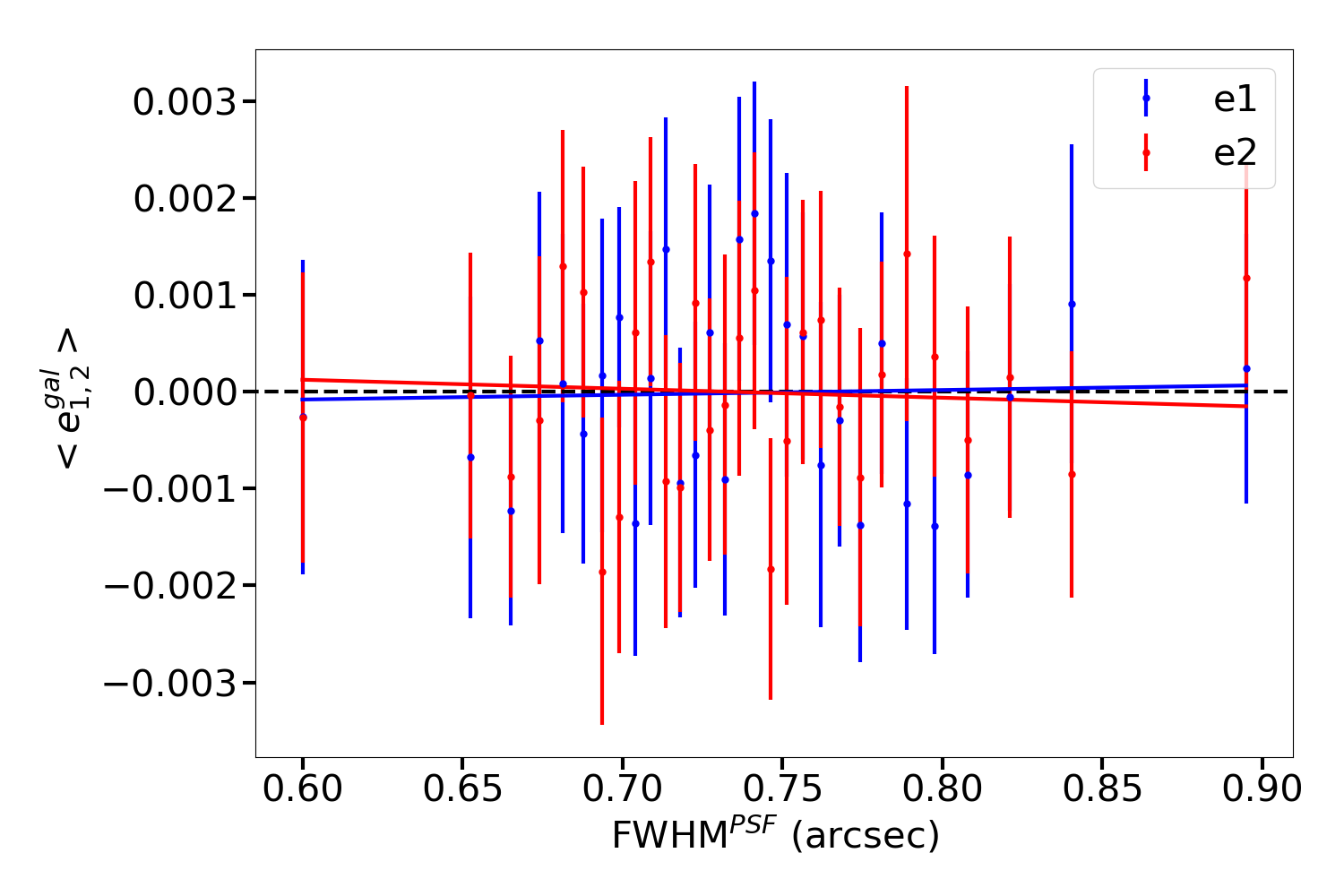}
        	\caption{PSF leakage using the averaged galaxy shape in bins of PSF size. The slopes are $\langle e_{1}^{\mathrm{gal}} \rangle = (0.0004 \pm 0.004)~\mathrm{FWHM}^{\mathrm{PSF}}$ and $\langle e_{2}^{\mathrm{gal}} \rangle = (-0.001 \pm 0.004)~\mathrm{FWHM}^{\mathrm{PSF}}$.}
        	\label{fig:PSF_leakage_simu_size}
        \end{figure}

    We present here the PSF leakage measured on the simulated data. In this particular case the PSF used during the shape measurement is the true model used to create the simulated images. This allows us to quantify the leakage from the shape measurement only, disregarding potential errors due to the modeling as we likely have on real data. In Fig.~\ref{fig:PSF_leakage_simu_ell} we do not observe any measurable dependencies between the PSF and galaxy ellipticities. The same conclusion is obtained with the PSF size as shown in Fig.~\ref{fig:PSF_leakage_simu_size}. These results reinforce our hypothesis that the leakage observed in the real data are coming mainly from the errors in the PSF model.

\section{Convergence maps}
\label{sec:annexe_mass_map}
\begin{figure*}
            \centering
            \hspace{-18.0mm}
            \includegraphics[width=1\textwidth]{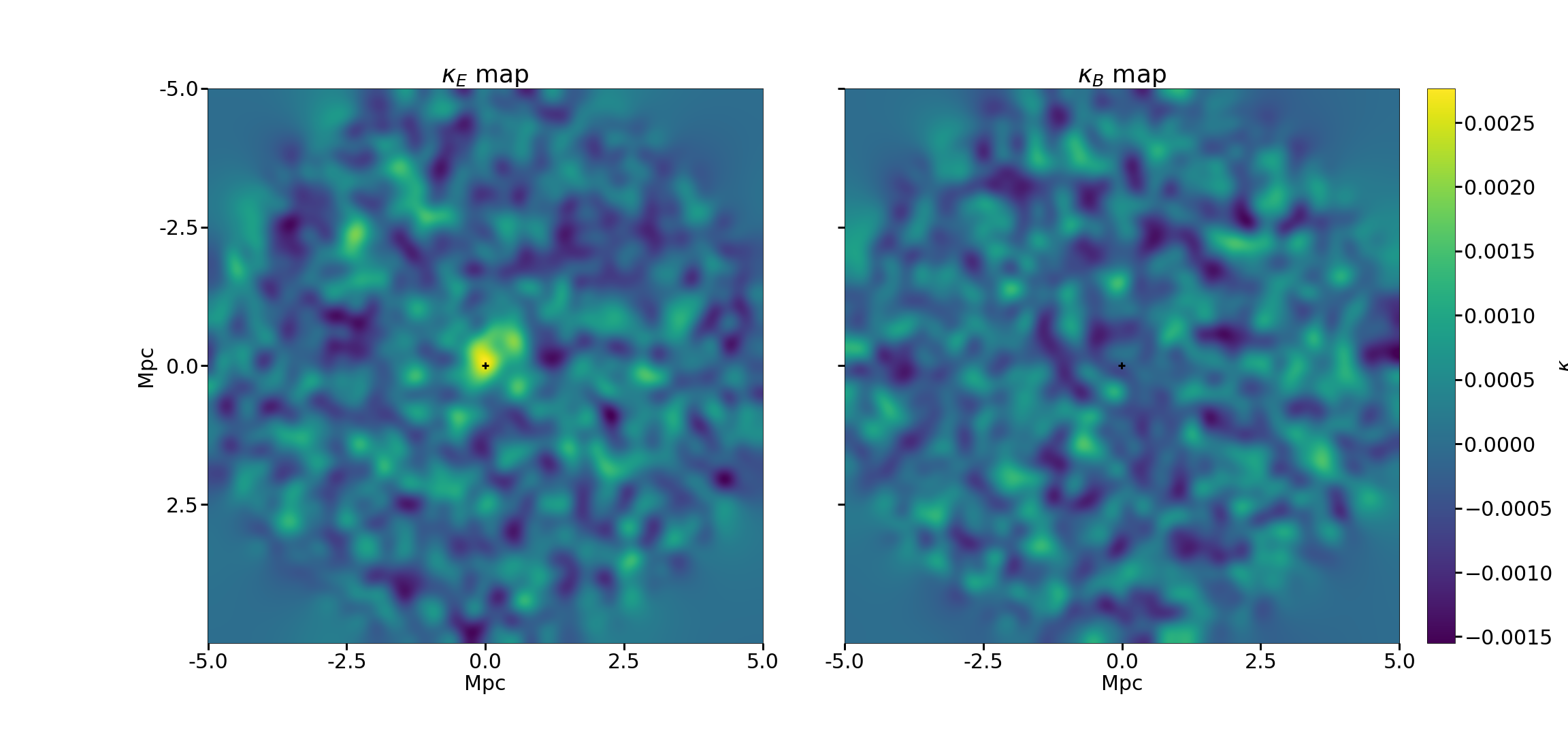}
            \caption{Mass maps stacked on the Planck cluster positions for P2. The galaxies for the tangential shear stacks are selected in a radius of $5~\mathrm{Mpc}$ around each cluster, where this distance is computed at the cluster redshift.}
            \label{fig:stacked_mass_map_P2}
        \end{figure*}

\begin{figure*}
            \centering
            \hspace{-18.0mm}
            \includegraphics[width=1\textwidth]{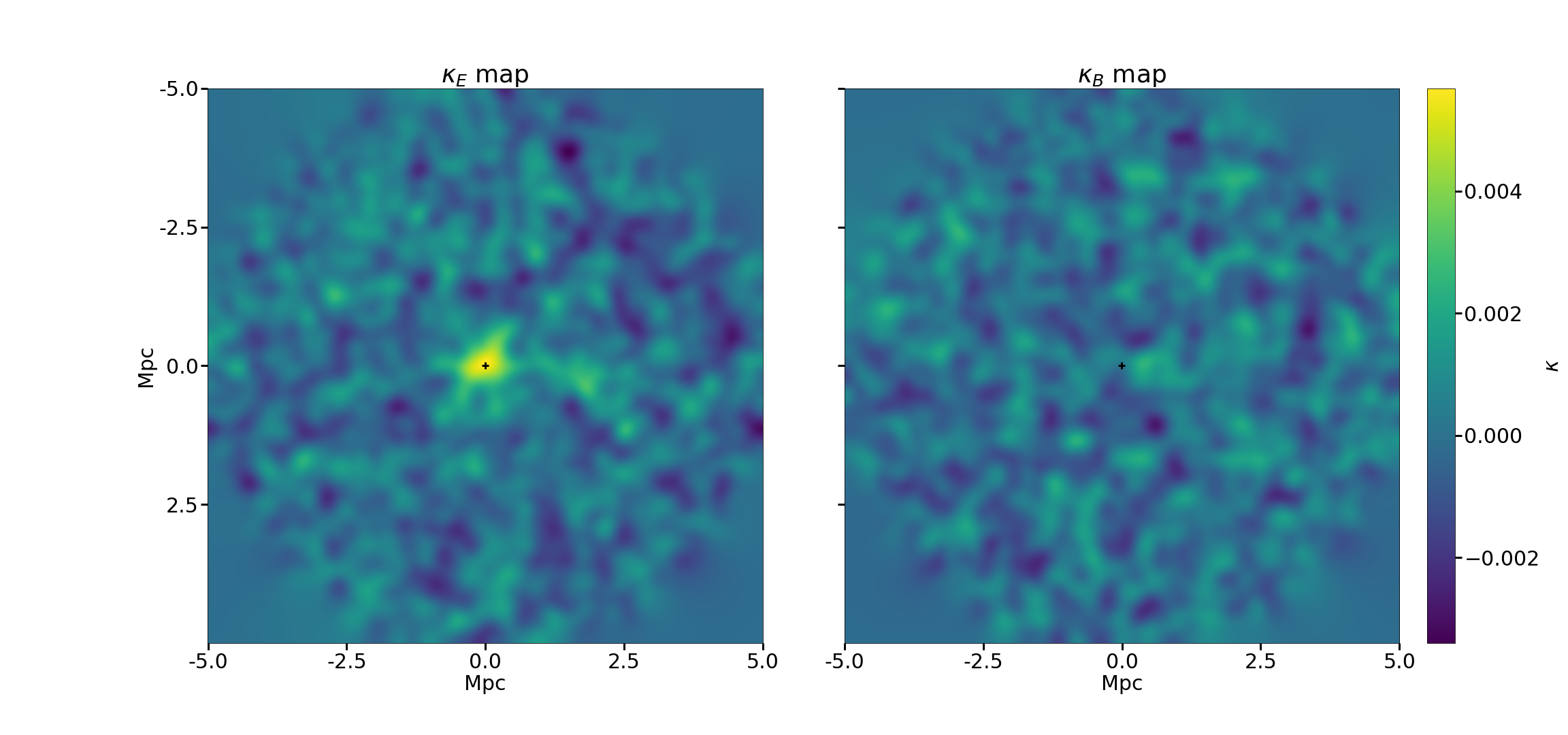}
            \caption{Mass maps stacked on the Planck cluster positions for P3. The galaxies for the tangential shear stacks are selected in a radius of $5~\mathrm{Mpc}$ around each cluster, where this distance is computed at the cluster redshift.}
            \label{fig:stacked_mass_map_P3}
        \end{figure*}

\begin{figure*}
            \centering
            \hspace{-18.0mm}
            \includegraphics[width=1\textwidth]{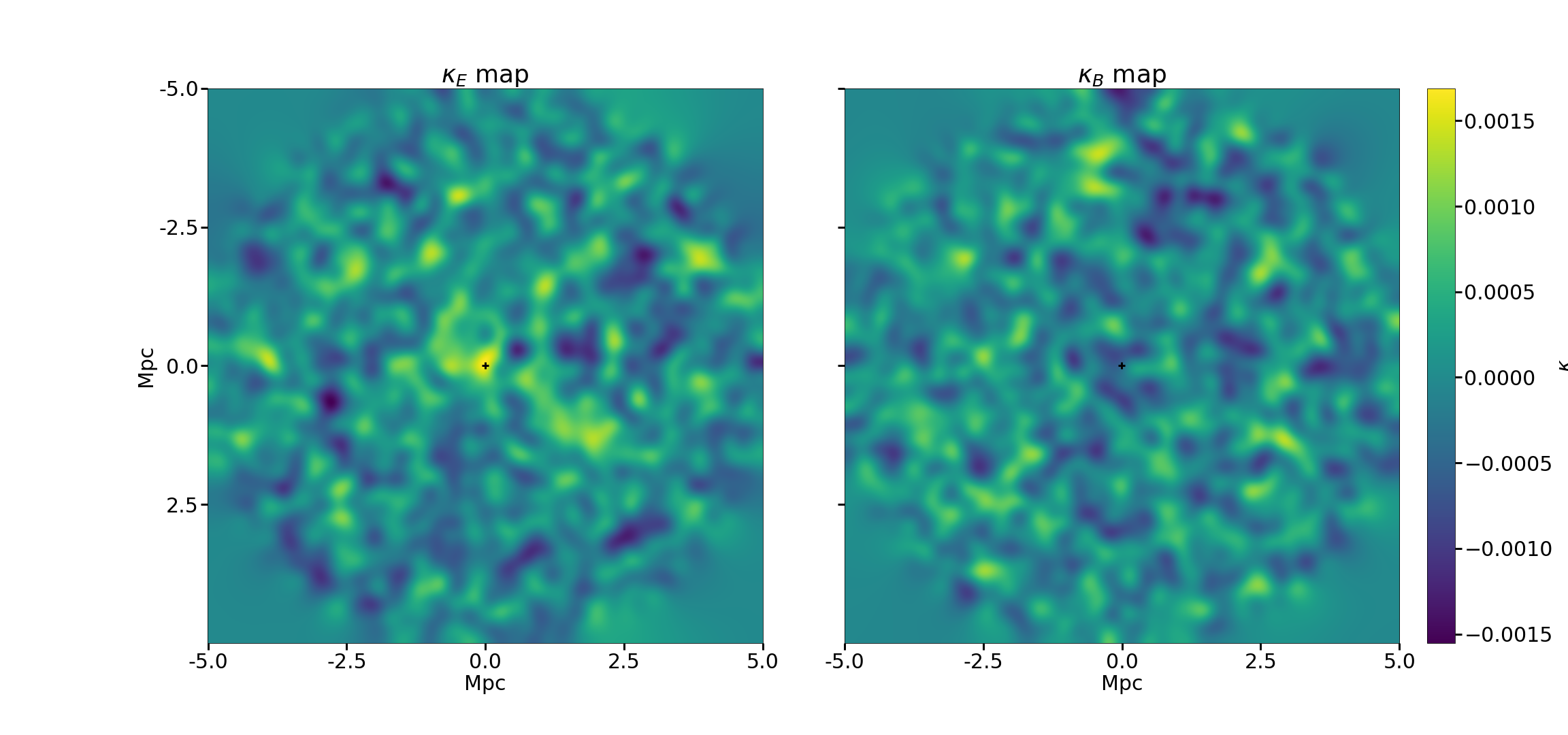}
            \caption{Mass maps stacked on the Planck cluster positions for P4. The galaxies for the tangential shear stacks are selected in a radius of $5~\mathrm{Mpc}$ around each cluster, where this distance is computed at the cluster redshift.}
            \label{fig:stacked_mass_map_P4}
        \end{figure*}

    In this section we show the convergence maps obtained through the method detailed in Sect.~\ref{sec:mass_map} for the 3 other patches that have been processed. At some cluster positions it is difficult to see a correlation with an over-density on the projected convergence maps. Yet, the stacked profiles shown in Figs.~\ref{fig:stacked_mass_map_P2}, \ref{fig:stacked_mass_map_P3} and \ref{fig:stacked_mass_map_P4} all shows a very clear signal on the E-mode while only noise is present on B-mode. This analysis tend to confirm the ability of our pipeline to accurately measure shapes with a low level of systematics.
    
    
        \begin{figure*}
    		\centering
    		\begin{minipage}{\textwidth}
    			\centering
    			\includegraphics[width=0.8\textwidth]{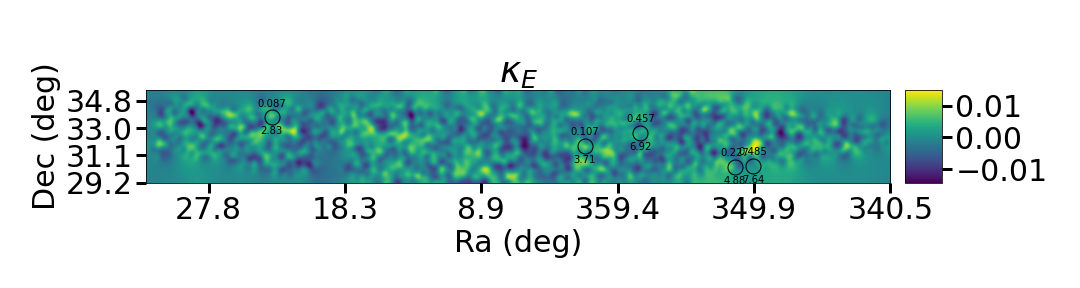}
    		\end{minipage}
    		
    		\begin{minipage}{\textwidth}
    			\centering 
    			\includegraphics[width=0.8\textwidth]{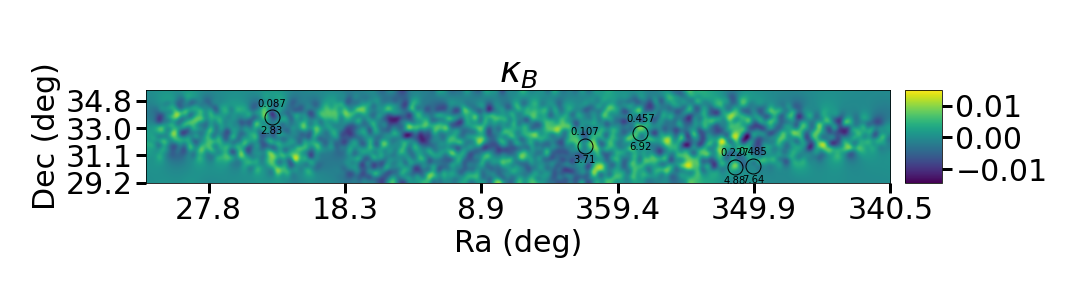}
    		\end{minipage}
    
    		\caption{Mass map for the patch P2. The \textit{black circles} represent the positions of Planck clusters. The value on \textit{top} of each cross is the cluster redshift, and the \textit{bottom} value indicates the SZ cluster mass ($10^{14}\mathrm{M}_{\odot}$). The \textit{top} (\textit{bottom}) panel shows the E-mode (B-mode).}
    		\label{fig:mass_map_P2}
    	\end{figure*}
        	
        \begin{figure*}
    		\centering
    		\begin{minipage}{\textwidth}
    			\centering
    			\includegraphics[width=0.8\textwidth]{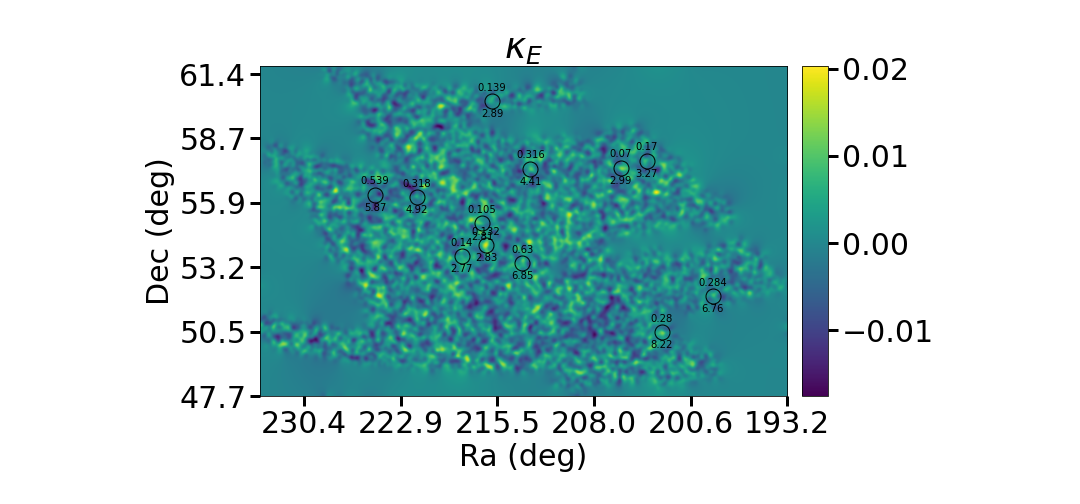}
    		\end{minipage}
    		
    		\begin{minipage}{\textwidth}
    			\centering 
    			\includegraphics[width=0.8\textwidth]{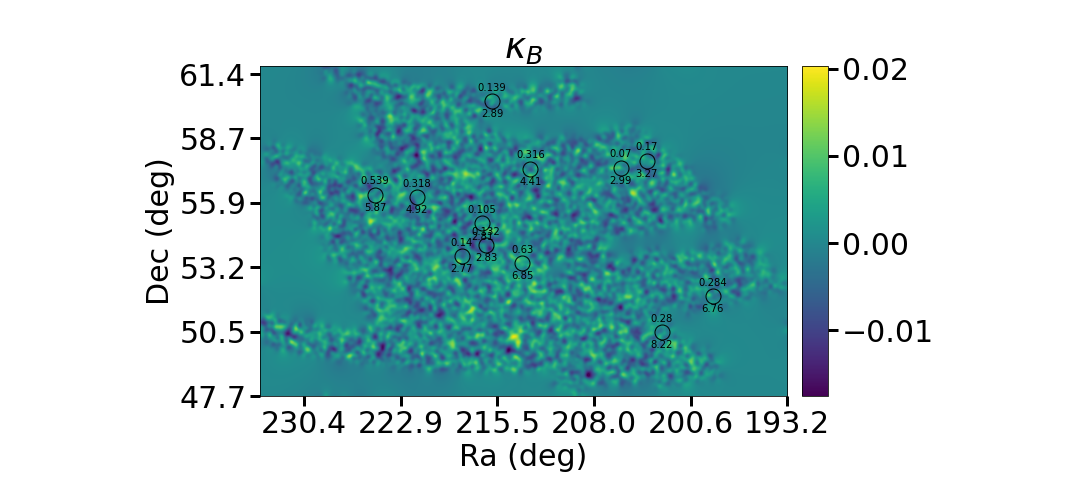}
    		\end{minipage}
    
    		\caption{Mass map for the patch P3. The \textit{black circles} represent the positions of Planck clusters. The value on \textit{top} of each cross is the cluster redshift, and the \textit{bottom} value indicates the SZ cluster mass ($10^{14}\mathrm{M}_{\odot}$). The \textit{top} (\textit{bottom}) panel shows the E-mode (B-mode).}
    		\label{fig:mass_map_P3}
    	\end{figure*}
        	
        \begin{figure*}
    		\centering
    		\begin{minipage}{\textwidth}
    			\centering
    			\includegraphics[width=0.8\textwidth]{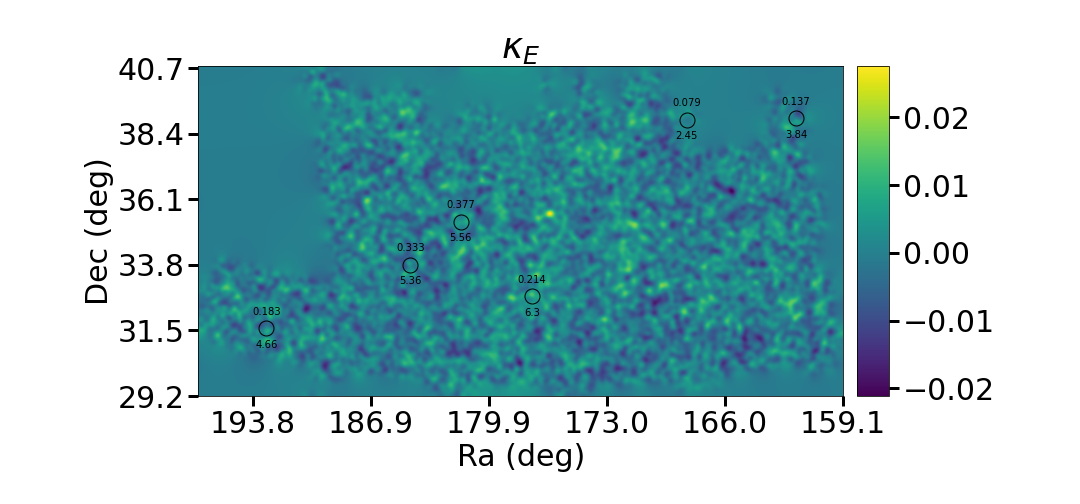}
    		\end{minipage}
    		
    		\begin{minipage}{\textwidth}
    			\centering 
    			\includegraphics[width=0.8\textwidth]{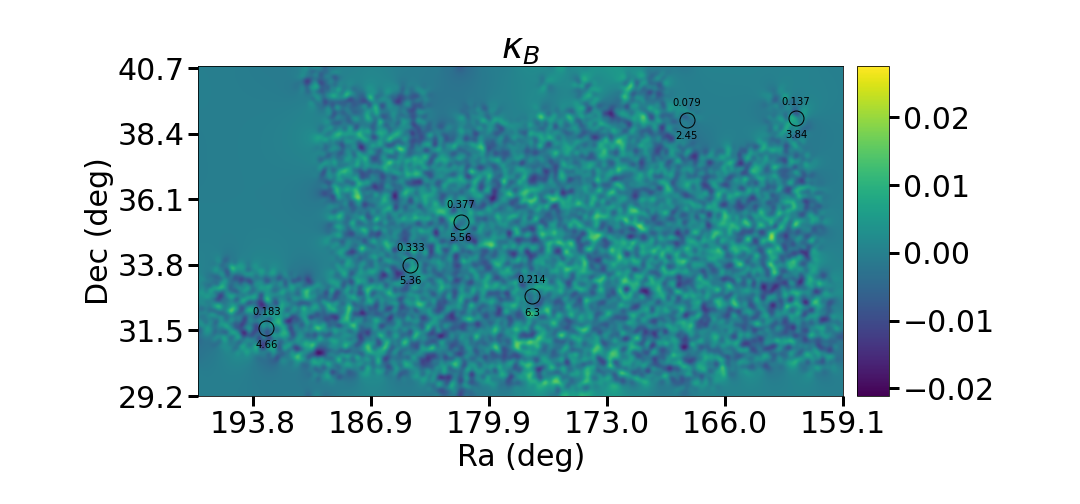}
    		\end{minipage}
    
    		\caption{Mass map for the patch P4. The \textit{black circles} represent the positions of Planck clusters. The value on \textit{top} of each cross is the cluster redshift, and the \textit{bottom} value indicates the SZ cluster mass ($10^{14}\mathrm{M}_{\odot}$). The \textit{top} (\textit{bottom}) panel shows the E-mode (B-mode).}
    		\label{fig:mass_map_P4}
    	\end{figure*}
        	
\end{appendix}


\end{document}